\documentclass[aps, prd, a4paper, twocolumn, showpacs, floatfix, superscriptaddress, longbibliography, nofootinbib]{revtex4-2}

\usepackage{amsmath}
\usepackage{amssymb}
\usepackage{amsfonts}
\usepackage{graphicx}
\usepackage{dcolumn}
\usepackage{bm}
\usepackage{hyperref}
\usepackage{amsopn}
\usepackage{mathtools}

\usepackage[mathlines]{lineno}

\newcommand{\dd}{\mathrm{d}}
\newcommand{\lie}{\pounds}
\newcommand{\del}{\partial}
\newcommand{\dirac}[2]{\delta^{#1}(#2)}
\newcommand{\Poisson}[2]{\left\{#1,\;#2\right\}}
\newcommand{\Comm}[2]{\left[#1,\;#2\right]}

\newcommand{\Prod}[2]{\left(#1, #2\right)}

\newcommand{\ci}{\mathsf{i}}
\newcommand{\Prp}{\mathsf{P}}

\newcommand{\CS}{\mathsf{J}}

\newcommand{\myfrac}[2]{%
    \setbox0\hbox{$#1$}        
    \dimen0=\wd0               
    \setbox1\hbox{$#2$}        
    \dimen1=\wd1               
    \ifdim\wd0<\wd1            
        \dfrac{#1\hfill}{#2}   
    \else
        \dfrac{#1}{#2\hfill}   
    \fi
}

\newcommand{\n}{n}

\newcommand{\ddt}{\dd{}t}
\newcommand{\ddx}{\dd^3x}

\newcommand{\ST}{\Sigma}

\newcommand{\mf}{\phi}
\newcommand{\pmf}{\Pi_\mf}

\newcommand{\mvo}{\hat{\chi}}
\newcommand{\fsmvo}{\tilde{\chi}}

\newcommand{\gH}{H}
\newcommand{\scd}{D}
\newcommand{\HF}[1]{\mathcal{Y}_{#1}}
\newcommand{\EV}[1]{\lambda_{#1}}

\newcommand{\HFi}{\bm{k}}

\newcommand{\BVi}{\mathsf{v}}
\newcommand{\BR}{\mathsf{R}}

\newcommand{\BRi}{\mathsf{r}}
\newcommand{\BS}{\mathsf{Q}}
\newcommand{\BSi}{\mathsf{q}}

\newcommand{\dBR}{\dot{\mathsf{R}}}

\newcommand{\AO}{\mathsf{a}}

\newcommand{\ket}[1]{\left\vert{#1}\right\rangle}

\newcommand{\braketOP}[3]{\left\langle#1\middle\vert#2\middle\vert#3\right\rangle}

\newcommand{\SM}{S}

\newcommand{\lapse}{\mathcal{N}}
\newcommand{\modesumele}{\dd\nu_{\HFi}}
\newcommand{\modesum}{\int\!\!\modesumele}
\newcommand{\modedelta}[2]{\dirac{3}{#1,#2}}

\DeclareFontFamily{U}{mathx}{\hyphenchar\font45}
\DeclareFontShape{U}{mathx}{m}{n}{<-> mathx10}{}
\DeclareSymbolFont{mathx}{U}{mathx}{m}{n}
\DeclareMathAccent{\widebar}{0}{mathx}{"73}

\newcommand{\cs}{c_s}

\newcommand{\vm}{M}
\newcommand{\vq}{Q}
\newcommand{\vn}{N}
\newcommand{\vv}{V}
\newcommand{\vu}{U}
\newcommand{\vg}{G}
\newcommand{\vl}{L}

\newcommand{\vI}{I}
\newcommand{\vp}{P}
\newcommand{\va}{A}
\newcommand{\vb}{B}

\newcommand{\PSV}{\mathbb{V}_{\HFi,t_0}}
\newcommand{\PSVr}{\PSV^r}
\newcommand{\PSVM}{\PSV^M}

\newcommand{\bigO}[1]{\mathcal{O}\!\left(#1\right)}

\DeclareMathOperator{\acosh}{\cosh^{-1}}
\DeclareMathOperator{\sh}{\sinh}
\DeclareMathOperator{\ch}{\cosh}
\DeclareMathOperator{\tah}{\tanh}

\begin{document}
\title{New formalism to define vacuum states for scalar fields in curved space-times}

\author{Mariana Penna-Lima}

\email{pennalima@unb.br}

\affiliation{Universidade de Bras\'{i}lia, Instituto de F\'{i}sica, Caixa Postal 04455, 70919-970, Bras\'{i}lia, DF, Brazil}

\author{Nelson Pinto-Neto}

\email{nelsonpn@cbpf.br}

\affiliation{COSMO -- Centro Brasileiro de Pesquisas
F\'{i}sicas, Xavier Sigaud, 150, Urca, 22290-180, Rio de Janeiro, Brasil}

\author{Sandro~D.~P.~Vitenti}
\email{vitenti@uel.br}
\affiliation{Departamento de F\'{i}sica, Universidade Estadual
de Londrina, Rod. Celso Garcia Cid, Km 380, 86057-970,
Londrina, Paran\'{a}, Brazil}

\date{\today}

\begin{abstract}
The problem of finding a vacuum definition for a single quantum field in curved space-times is discussed under a new geometrical perspective. The phase space dynamics of the quantum field modes are mapped to curves in a 2-dimensional hyperbolic metric space, in which distances between neighbor points are shown to be proportional to the Bogoliubov coefficients associated with their corresponding mode solutions in phase space. The vacuum state for each mode is then defined as the unique trajectory from which all mapped phase space solutions move within thin annular regions around it. This property implies the stability of the vacuum state: solutions evolved from a point in this trajectory stay close to it as both evolve, and the particle creation is therefore minimized. The new approach is applied to the well-known cases of the time-independent dynamics, where the solutions draw circles around this curve, and when the adiabatic approximation is valid. The analysis is then extended to time-dependent cases in which the adiabatic approximation is not applicable, in the super-Hubble or low-frequency regimes. It is shown that stability trajectories can also be found in these situations, and stable quantum vacua can be obtained. This new formalism is applied to two situations: de Sitter space, where the Bunch-Davies vacuum is obtained in a completely different manner through an analysis in the super-Hubble regime, and in the context of cosmological bouncing models, in which the contracting phase is dominated by a cosmological constant in the asymptotic past. A new vacuum state for cosmological perturbations is proposed in this situation.
\end{abstract}


\maketitle

\section{Introduction} 

The determination of the vacuum state of quantum fields in curved space-times is a quite intricate subject, which has been studied for a long time \cite{Parker1968, Parker1969, DeWitt1975, Birrell1982, Fulling1989, Wald1994}. Either one has many non-equivalent possible candidates \cite{Fulling1973, Hawking1974, Boulware1975, Hawking1975, Wald1975, Unruh1976}, or none at all. This is a crucial problem for theoretical physics, as long as space-time is indeed curved because of gravity, which is universal. Usually, one appeals to the ultra-violet limit, restricting the analysis to a large enough ensemble of field modes with mode wavelengths much smaller than the curvature scale of the space-time region with physical interest. In this situation, the curvature of space-time is almost irrelevant, one is close to the flat Minkowski case, where the vacuum state is well defined, and an adiabatic approximation is available to guarantee the stability of the chosen vacuum for some finite time interval. In the case of de Sitter space, i.e., a curved space in which the curvature scale (given by $\Lambda^{-1/2}$, where $\Lambda$ is the cosmological constant) is globally constant in space and time, the adiabatic approximations is one of the methods used to define the globally stable Bunch-Davies vacuum. Note, however, that this is one of the possible quantum vacua in de Sitter space, see Ref.~\cite{Allen1985} for a classification and properties of the possibilities. Furthermore, in inflationary models there is also the scenario where one must impose initial conditions at a finite time in the past where perturbations wave-lengths are not completely sub-horizon. In these situations the ambiguity in the vacuum choice can lead to observable imprints in the Cosmic Microwave Background (CMB)~\cite{Handley2016, Agocs2020, GesseyJones2021}.

This paper aims to investigate whether one can define the equivalent of an adiabatic vacuum in a regime where the frequencies of the field modes are irrelevant to their dynamics, or their mode wavelengths are much bigger than the curvature scale of space-time, the reverse of the usual approach. After reviewing the procedure of choosing a complete set of solutions for the field variables in phase space to define the creation/annihilation operators set, we identify the minimum number of degrees of freedom necessary to define a vacuum, which are then mapped to a 2-dimensional hyperbolic space $\mathbb{H}^2$ in which distances between points can be defined. In this space, canonical transformations are seen as boosts and rotations. We then analyze the usual vacuum prescriptions for time-independent Hamiltonians, and time-dependent situations where the adiabatic approximation is attainable and an adiabatic vacuum can be defined. In the first case, we show that the usual vacuum choice corresponds to a single point in $\mathbb{H}^2$ from which all mapped phase space solutions move in circles around it. In the second case, after making appropriate time-dependent boosts and rotations, the adiabatic vacua correspond to points in a small region of $\mathbb{H}^2$ which is the central neighborhood of a thin annular region to which all solutions of the mode equations are mapped. Increasing the order of the adiabatic approximation, the thinner becomes this annular region. These properties are equivalent to the stability of the vacuum choice, being exactly it in the first case, while approximately in the second one.

To extend this analysis to time-dependent cases in which the adiabatic approximation is not valid, we investigate in detail physical situations in which the mode frequencies are irrelevant to the dynamics of the modes. Again, after performing convenient time-dependent boosts and rotations in $\mathbb{H}^2$, we find in the new frame stability points with similar properties to the known previous cases. Such points yield stable quantum states, in the sense that particle production is null up to some arbitrary order of approximation in the finite time interval considered, yielding, as before, a sensible quantum vacuum choice. Note that, in both time-dependent cases (adiabatic and non-adiabatic), these special points are transformed into small curves in $\mathbb{H}^2$ when we get back to the original frame in which these problems are usually formulated, by performing the inverse time-dependent boosts and rotations that have been implemented.

The new formalism is then applied to three important physical situations. As we know, a positive cosmological constant is by far the simplest explanation for the present acceleration of the Universe, and many other refined cosmological observations, for instance, see \cite{Planck2018}. Hence, it is not by chance that the standard cosmological model is called the $\Lambda$CDM model. In the case of bouncing models in which the cosmological constant is not considered, the universe tends to be flat in the far past of the contracting phase, allowing the prescription of initial conditions in terms of an adiabatic quantum vacuum, which can be easily defined in this situation. 
Calculations assuming this approach concludes that one can obtain almost scale-invariant cosmological perturbations if the contracting phase is dominated by an almost pressureless fluid, perhaps dark matter~\cite{Wands1999,Finelli2002, Peter2002}. We get the same results using the new formalism developed in the present paper. 
However, if a positive cosmological constant is present, the asymptotic past of bouncing models will approach de Sitter rather than Minkowski spacetime. In this asymptotic limit, the physical mode frequencies squared become negligible concerning the cosmological constant $\Lambda$, or equivalently, the mode wavelengths become much bigger than the Hubble radius given by $\Lambda$. Thus, the adiabatic approximation is not valid in this asymptotic region. 
Note that there is a period in the cosmic evolution of such models when the frequencies associated with large cosmological wavelengths become relevant and dominate the evolution of the perturbation field, allowing the usual prescription of an adiabatic vacuum for them. Unfortunately, this period is short, and
stability is not guaranteed in the far past when the cosmological constant dominates over the pressureless fluid.
This problem has been studied and discussed in Ref.~\cite{Maier2011}.\footnote{In the case of quintessence models for dark energy, this problem can be overcome~\cite{Bacalhau2018}.} Nevertheless, using our new method, we are able to find stability curves in $\mathbb{H}^2$ to which one can associate a stable quantum vacuum state in the asymptotic past when the cosmological constant dominates. This is a new quantum vacuum state which naturally appears in this class of models, whose physical consequences can now be explored.

Finally, we use the new formalism to re-obtain the Bunch-Davies vacuum of de Sitter space in a complete different manner by looking at the super-Hubble regime, the time period in which the physical wavelengths of the modes are much bigger than the Hubble radius of de Sitter space, and when the adiabatic approximation is not applicable. Surprisingly, after implementing some boosts and rotations, we find one stability point in $\mathbb{H}^2$ with exactly the same properties of the stability point we find in the case of time-independent Hamiltonians in flat space. This stability point is shown to be the Bunch-Davies vacuum. In the case of a massless scalar field, we obtain this result without ever using the general solution of the mode equations.

The paper is divided as follows: in Section~\ref{sec:quant:gen} we summarize the formalism which we will use to construct the quantities necessary to define a vacuum state, and the new representations involving the $\mathbb{H}^2$ space. In Sec.~\ref{VDC} the new formalism is applied to the well-known and simple case of a time-independent Hamiltonian, and in Sec.~\ref{sec:gcase} to the time-dependent case. We organize this last in two subsections. In the first one, Sec.~\ref{subsec:nu}, we recover the usual adiabatic prescription to obtain a quantum vacuum when the mode frequencies dominate the dynamics of the field modes, including the well-known results.
The procedure is equivalent to many other implementations, but is simpler and more adequate for numerical calculations. In the second subsection, Sec.~\ref{subsec:V}, we apply the formalism to the case where the mode frequencies are irrelevant for the dynamics, and the adiabatic approximation is not valid, obtaining stable vacuum states as well. In Sec.~\ref{sec:Applications} we apply the new formalism to the three cases mentioned above: matter bounce without and with a cosmological constant, and de Sitter space, with the new method to obtain the Bunch-Davies vacuum. We end up with the conclusions in Sec.~\ref{sec:conclusion}. There are three appendixes: Appendix~\ref{app:PSR} shows the details of the constructions, spaces, and representations described in Sec.~\ref{sec:quant:gen}, App.~\ref{app:HT} presents the Hamiltonian tensor in the new language, and App.~\ref{app:adiab} shows the construction details of the adiabatic frames used in Sec.~\ref{subsec:nu}, and their associated propagators.   

\section{Field Quantization and Bogoliubov Coefficients}
\label{sec:quant:gen}

In this section we summarize the field quantization procedure using three different representations: phase space vectors, $2\times 2$ real matrices, and Clifford algebra. The aim is to pave the way to fix the vacuum state using a minimal set of variables. 

We first focus on the symplectic structure behind the quantization of a single field degree of freedom and its momentum described by quadratic Hamiltonians. It contains the essence of our method, besides complying with many cases relevant to curved spacetimes and Cosmology. We use a 3+1 split $(x, t)$ of spacetime, with globally defined spatial hypersurfaces, where $x$ represents the three-dimensional space coordinates and $t$ the time. In Refs.~\cite{Vitenti2015, Peter2016} this procedure is generalized to an arbitrary number of degrees of freedom.

We define the phase vector field $\chi_a$ and the symmetric Hamiltonian tensor $\gH^{ab}$, respectively, as
\begin{equation}\label{eq1}
\chi_a \doteq (\mf,\;\pmf), \qquad\gH(\chi) = \frac{1}{2}\chi_a\gH^{ab}\chi_b,
\end{equation}
where $\mf$ represents a generic scalar field, and $\pmf$ its canonical
momentum.\footnote{In this work we use the symbol $\doteq$ to define components
	of vectors and tensors, and $\equiv$ for general definitions.} Then, it is easy
to see that the action can be written as
\begin{equation}\label{eq:action:new}
\begin{split}
\mathcal{S} (\mf,\pmf) &= \frac{1}{2}\int\ddx\,\ddt\left(\ci\chi_a\SM^{ab}\dot{\chi}_b - \chi_a\gH^{ab}\chi_b\right),
\end{split}
\end{equation}
where the symplectic matrix and its inverse are defined by
\begin{equation}\label{eq:def:SM}
\SM_{ab} \doteq \ci\left( \begin{array}{cc}
0 & 1 \\
-1 & 0 \end{array} \right),\quad \SM^{ab} \doteq \ci\left( \begin{array}{cc}
0 & 1 \\
-1 & 0 \end{array} \right),
\end{equation}
in which the imaginary unit $\ci$ is added for later convenience.

Extremizing the action~\eqref{eq:action:new} with respect to $\mf$ and $\pmf$, we find the equations of motion
satisfied by the phase vector field, namely
\begin{equation}
\label{hamilton:equations}
\ci{\dot{\chi}_a} = \SM_{ab}\gH^{bc}\chi_c,
\end{equation}
where the over-dot represent differentiation with respect to the time $t$.

Within this mathematical structure, the Poisson bracket of any two functionals
$F_1(\mf, \pmf)$ and $F_2(\mf, \pmf)$ can be written as
\begin{equation}
\Poisson{F_1}{F_2} = -\ci\int\limits_\ST\ddx\myfrac{\delta F_1}{\delta\chi_a(x)}\SM_{ab}\myfrac{\delta F_2}{\delta\chi_b(x)},
\end{equation}
where $\ST$ represents one particular spatial hypersurface of the $3+1$
splitting. Using the definitions above, we have
\begin{equation}
\label{poisson}
\begin{split}
\Poisson{\chi_a(x_1)}{\chi_b(x_2)} &= -\ci\SM_{ab}\dirac{3}{x_1-x_2}.
\end{split}
\end{equation}

Applying the canonical quantization rules, the classical fields are promoted to
Hermitian operators, $\hat{\mf}$ and $\hat{\Pi}_{\mf}$, and the Poisson brackets
Eq.~\eqref{poisson} lead to the equal-time commutation relations
\begin{equation}\label{eq:def:comm:chi1:chi2}
\Comm{\mvo_a(x_1)}{\mvo_b(x_2)} = \SM_{ab}\dirac{3}{x_1 - x_2},
\end{equation}
which decompose into the familiar commutators
\begin{align}
\Comm{\hat{\mf}(x_1)}{\hat{\Pi}_{\mf}(x_2)} &= \ci\dirac{3}{x_1-x_2}, \\
\Comm{\hat{\mf}(x_1)}{\hat{\mf}(x_2)} &= 0 = \Comm{\hat{\Pi}_{\mf}(x_1)}{\hat{\Pi}_{\mf}(x_2)}.
\end{align}
From here on we set $\hbar = 1$.

Canonical quantization defines the operator algebra, but does not provide a route to build their representations. In Quantum Mechanics, where the number of degrees of freedom is finite, the lack of a natural procedure for constructing the representations is irrelevant, because all representations are unitarily equivalent to each other due to the Stone--von Neumann theorem~\cite{Reed1979}. In the case of field quantization, however, we have an infinite number of degrees of freedom, and the Stone--von Neumann theorem does not apply. Consequently, for fields, canonical quantization no longer yields a complete description of the quantum system, and one must do it by choosing a class of unitary equivalent representations.

The general procedure to obtain a representation starts by complexifying the space of field solutions of the second-order differential equations arising from the action \eqref{eq:action:new}, which we denote by $\{\phi_{\HFi}(t,x)\}$, where $\HFi$ represents all the necessary indices labeling a complete set of solutions\footnote{For example, for flat spatial sections we can decompose the function in Fourier space and, in this case, $\HFi$ would be the Fourier mode vector.} allowing complex initial conditions. Each $\phi_{\HFi}(t,x)$ with a particular label $\HFi$ is called a mode.

It is usually hard to obtain a set of analytic solutions of the second-order equations of motion for $\{\phi_{\HFi}(t,x)\}$. However, there is a one-to-one mapping between the solutions $\{\phi_{\HFi}(t,x)\}$ defined on the whole manifold, and the fields $\left(\phi_{\HFi}(t_0, x),\;\Pi_{\phi_{\HFi}}(t_0, x)\right)$ defined in phase space at a specific time slice (uniqueness of solutions). Hence, instead of working with a set of solutions $\{\phi_{\HFi}(t,x)\}$, we introduce the set of functions $\{\chi_{\HFi,a}(t_0,x)\}$, where
\begin{equation}
\chi_{\HFi,a}(t_0, x) \doteq \left(\phi_{\HFi}(t_0, x),\;\Pi_{\phi_{\HFi}}(t_0, x)\right)
\end{equation}
describes the initial conditions for each mode, yielding a unique phase space solution arising from Eq.~\eqref{hamilton:equations}. One can choose these functions arbitrarily without any knowledge of the solutions of the equations of motion (we are not considering here constrained Hamiltonian systems).

The inner product between two different elements of $\{\chi_{\HFi,a}(t_0,x)\}$, $\BR_{\HFi_1,a}(t_0,x)$ and $\BR_{\HFi_2,a}(t_0,x)$, omitting the $(t_0,x)$ dependence, reads
\begin{equation}
\label{eq:prod:BF1:BF2}
\Prod{\BR_{\HFi_1}}{\BR_{\HFi_2}} = \int\limits_{\ST(t_0)}\dd^3x \BR_{\HFi_1,a}^*\SM^{ab}\BR_{\HFi_2,b}.
\end{equation}
The phase space set of vectors $\{\chi_{\HFi,a}(t_0,x)\}$ is said to be
complete if they satisfy,
\begin{align}\label{eq:def:prod:BF}
&\Prod{\BR_{\HFi_1}}{\BR_{\HFi_2}} = \modedelta{\HFi_1}{\HFi_2}, \qquad \Prod{\BR_{\HFi_1}^*}{\BR_{\HFi_2}} = 0, \\
&\modesum\left[\BR_{\HFi,a}\Prod{\BR_{\HFi}}{f} - \BR_{\HFi,a}^*\Prod{\BR_{\HFi}^*}{f}\right] = f_a,
\end{align}
where $f$ is an arbitrary phase space vector and $\dd\nu_{\bf{k}}$ is the measure of the solution's space. Then, one can express the phase space field operator $\mvo_a$ as
\begin{equation}
\label{field:operator:expansion}
\mvo_a = \modesum\left(\BR_{\HFi,a}\AO_{\HFi} + \BR_{\HFi,a}^*\AO_{\HFi}^\dagger\right),
\end{equation}
where
\begin{equation}
\label{creation:operator}
\AO_{\HFi} \equiv \Prod{\BR_{\HFi}}{\mvo}, \qquad \AO^\dagger_{\HFi} \equiv -\Prod{\BR_{\HFi}^*}{\mvo}
\end{equation}
are the so-called annihilation and creation operators, respectively. The field commutation relations \eqref{eq:def:comm:chi1:chi2} and Eq.~\eqref{eq:def:prod:BF} lead to
\begin{equation}\label{eq:comm:a:ad}
\Comm{\AO_{\HFi_1}}{\AO^\dagger_{\HFi_2}} = \modedelta{\HFi_1}{\HFi_2},\qquad \Comm{\AO_{\HFi_1}}{\AO_{\HFi_2}} = 0,
\end{equation}
which allows to define the Fock space as the space composed by the vacuum state, which satisfies $\AO_{\HFi}\ket{0} = 0$ for all modes $\HFi$, and all other states obtained through a finite number of applications of the operator $\AO^\dagger_{\HFi}$ with different indices $\HFi$ in $\ket{0}$.

One reason to choose the inner product \eqref{eq:def:prod:BF} in order to define the completeness of solutions is because it satisfies the property
\begin{equation}\label{eq:cons:prod}
\begin{split}
\ci\lie_\n\Prod{\BR_{\HFi_1}(t)}{\BR_{\HFi_2}(t)} &=  0,
\end{split}
\end{equation}
where $\lie_\n$ denotes the Lie derivative in the direction of the normal of the hypersurfaces $t=$ const., and $(\BR_{\HFi_1}(t), \BR_{\HFi_2}(t))$ are the phase space vectors time evolved respectively from $(\BR_{\HFi_1}(t_0), \BR_{\HFi_2}(t_0))$ according to the Hamilton equations \eqref{hamilton:equations},
\begin{equation}\label{eq:motion:BF}
\ci\dot{\BR}_{\HFi,a}(t) = \SM_{ab}\myfrac{\del\gH}{\del \BR_{\HFi,b}} = \SM_{ab}\gH^{bc}\BR_{\HFi,c}(t).
\end{equation}
Hence, if the set $\{\chi_{\HFi}(t_0)\}$ is complete in phase space at $t_0$, then $\{\chi_{\HFi}(t)\}$ is complete in phase space at any $t$ as long as all modes satisfy the Hamilton equations \eqref{eq:motion:BF}. These results are valid for any quadratic Hamiltonian system where $\SM_{ab}\gH^{bc}$ is self-adjoint with respect to the product~\eqref{eq:prod:BF1:BF2}.

In summarizing, given a complete set of phase space functions $\BR_{\HFi,a}$ defined on a spatial slice, we can decompose the quantum field in terms of creation and annihilation operators and obtain the Fock space representation naturally defined by them.


The problem of finding a complete set of functions $R_{\HFi,a}(t_0,x)$ (and
consequently $R_{\HFi,a}(t,x)$) satisfying Eq.~\eqref{eq:def:prod:BF} can be
simplified by focusing on functions of the form
\begin{equation}\label{eq:def:BF:split}
R_{\HFi,a}(t_0,x) = \BRi_{\HFi,a}(t_0)\HF{\HFi}(x),
\end{equation}
where $\BRi_{\HFi,a}(t_0)$ are arbitrary complex constants and $\HF{\HFi}(x)$ are the usual Laplacian eigenfunctions defined as
\begin{equation}\label{eq:laplace}
	\begin{split}
&\tilde{\scd}^2\HF{\HFi}(x) = -\EV{\HFi}^2\HF{\HFi}(x),\\
 &\int\limits_\ST\dd^3x \HF{\HFi_1}(x)\HF{\HFi_2}(x) = \modedelta{\HFi_1}{\HFi_2},
\end{split}
\end{equation}
with eigenvalues $-\EV{\HFi}^2$, where $\tilde{\scd}^2$ represents the Laplacian operator. For flat hypersurfaces, for example, we can choose $\HF{\HFi}$ as plane waves and, in this case, $\HFi$ is the mode vector and $\EV{\HFi}^2 = \HFi\cdot\HFi$.\footnote{In this work, we assume that $\HF{\HFi}$ are real functions, thus, instead of plane waves we should use the Hartley kernel.} Note that, for example, in a Friedmann geometry $\tilde{\scd}^2$ is the conformal Laplacian, since $\tilde{\scd}^2$ is constant, the eigenvalues $\EV{\HFi}^2$ are also constant.

Omitting again the $(x,t)$ dependencies, the product of two functions is given by
\begin{align}
\left.\Prod{\BR_{\HFi_1}}{\BR_{\HFi_2}}\right\vert_{t_0} &= \BRi_{\HFi_1a}^*\BRi_{\HFi_2b}\SM^{ab}\modedelta{\HFi_1}{\HFi_2},\\
\left.\Prod{\BR_{\HFi_1}^*}{\BR_{\HFi_2}}\right\vert_{t_0} &= \BRi_{\HFi_1a}\BRi_{\HFi_2b}\SM^{ab}\modedelta{\HFi_1}{\HFi_2} = 0.
\end{align}
The last equality comes from the fact that the $\delta$-function implies $\HFi_2 = \HFi_1$, and $\SM^{ab}$ is anti-symmetric. Hence, in order to have a normalized basis, we only need to impose
\begin{equation}
\label{normal:basis}
\BRi_{a}^*\BRi_{b}\SM^{ab} = 1,
\end{equation}
Consequently, using Eq.~\eqref{eq:def:BF:split} we turned the problem of finding a complete set of modes into the problem of finding a complete set of complex vectors $\{\BRi_{\HFi a}\}$, each one belonging to a bi-dimensional complex vector space $\PSV$ such that, for each mode $\HFi$, $\BRi_{a}^*\BRi_{b}\SM^{ab} = 1$ is satisfied. Using the (non-degenerated) symplectic matrix $\SM^{ab}$, we can map vectors into linear functionals $\BRi^{a} \equiv \SM^{ab}\BRi_{b}$\footnote{From now we will omit the label $\HFi$ for simplicity, introducing the $\HFi$ dependency explicitly only when necessary}. That is, we can use $\SM^{ab}$ and its inverse to ``raise'' and ``lower'' indices. In the App.~\ref{app:PSR} we develop all the necessary mathematical tools used to represent points in phase space.

\subsection{Bogoliubov coefficients and time evolution}
\label{sec:bogo}

As we have seen, expansion of the field operator as in Eq.~\eqref{field:operator:expansion} depends on the basis choices, leading to potentially different Fock spaces. Hence, the vacuum state may be inherently ambiguous (which is not the case when Poincar\'e invariance is required, and then all choices are unitarily equivalent), and we need a criterion to choose the basis. In the following, we quantify this ambiguity. 

Let us take $\BR_{\HFi}=\BRi\HF{\HFi}$. Any product in the form $\left.\Prod{\BRi\HF{\HFi}}{\mvo}\right\vert_{t_0}$ can be expressed as
\begin{equation}\label{eq:defaa}
\left.\Prod{\BRi\HF{\HFi}}{\mvo}\right\vert_{t_0} = \BRi_a^*\SM^{ab}\fsmvo_{\HFi ,b}(t_0) = \BRi^{a*}\fsmvo_{\HFi ,a}(t_0),
\end{equation}
where the operator $\fsmvo_{\HFi ,a}(t)$ reads
\begin{equation}\label{fsmvo}
\fsmvo_{\HFi,a}(t) \equiv \int_{\ST}\dd^3x \HF{\HFi}\mvo_a(t), \quad \mvo_a(t) = \modesum \HF{\HFi}\fsmvo_{\HFi,a}(t).
\end{equation}
Consequently, the annihilation
and creation operators, defined by the vector $\BRi_a$, are
\begin{equation}\label{eq:def:AO:fs}
\AO_{\HFi,\BRi} = \BRi^{a*}\fsmvo_{\HFi ,a}(t_0), \qquad \AO^\dagger_{\HFi,\BRi} = \BRi^{a}\fsmvo_{\HFi ,a}(t_0),
\end{equation}
which can be easily inverted using the projector in Eq.~\eqref{eq:def:complet} providing
\begin{equation}\label{eq:fsmvo:basis}
\fsmvo_{\HFi ,a}(t_0) = \BRi_a\AO_{\HFi,\BRi} + \BRi^*_a\AO^\dagger_{\HFi,\BRi}.
\end{equation}
As before, we will omit the label $\HFi$ in $\fsmvo_a(t)$ and $\AO$, writing it only when necessary.

A phase space vector $ \BRi_a$ at the hypersurface $t_0$ defines a
function in the whole time interval that satisfies
\begin{equation}\label{eqmovY}
\ci\dBR_a = \SM_{ab}\gH^{bc}\BR_c, \qquad \BR_a(t_0) = \BRi_a,
\end{equation}
where all Laplacian operators appearing in $\gH^{bc}$ must be substituted by
$-\EV{\HFi}^2$, so that $\BR_a$ is a function of time only.\footnote{The
	uppercase phase vectors $\BR_a$ are always considered functions of $t$ and we
	will write the time dependency only to avoid ambiguities.} This equation is a
consequence of the splitting given in Eq.~\eqref{eq:def:BF:split}, which was
possible because the equations of motion are separable in the time variable, as
are the cases of interest in this paper. We will denote the phase vector defined
in an initial hypersurface and the time-dependent phase vector that uses it as the
initial condition and satisfies Eq.~\eqref{eqmovY} with the same lowercase and
uppercase letters, respectively, for instance, $\BRi_a$ ($\BSi_a$) is the initial
condition for $\BR_a$ ($\BS_a$).

It is clear that the phase space operator $\fsmvo_a(t)$ also satisfies the same
Eq.~\eqref{eqmovY}. Therefore, as we already mentioned, $\AO_\BRi$ is constant
and consequently
\begin{equation}\label{aconst}
\AO_\BRi = \BRi^{a*}\fsmvo_a(t_0) = \BR^{a*}(t)\fsmvo_a(t),
\end{equation}
for any hypersurface $t$. Moreover, any product in the same form is constant and consequently
\begin{equation}\label{constproduct}
\BRi^{a*}\BSi_a = \BR^{a*}(t)\BS_a(t).
\end{equation}

Each normalized phase space vector basis defines a representation for the
quantization procedure. Two different basis, labeled respectively as $\BRi_a$
and $\BSi_a$, can lead to different representations (which can be
non-unitarily-equivalent). Assuming that $\BRi_a$ and $\BSi_a$ are two
normalized phase space vectors and using the projectors in
Eq.~\eqref{eq:def:complet}, we can write
\begin{equation}
\BRi_a = \alpha_{\BRi,\BSi}\BSi_a - \beta_{\BRi,\BSi}\BSi_a^*,
\end{equation}
where the products
\begin{equation}\label{eq:def:alpha:beta}
\alpha_{\BRi,\BSi} \equiv \BRi_a\BSi^{a*} = \BSi^*_a\SM^{ab}\BRi_b, \quad \beta_{\BRi,\BSi} \equiv -\BRi_a\BSi^a = \BSi_a\SM^{ab}\BRi_b,
\end{equation}
are constant due to Eq.~\eqref{constproduct} and satisfy $\alpha_{\BRi,\BRi} = 1$
and $\beta_{\BRi,\BRi} = 0$. Using the projectors it is also easy to show that
\begin{equation}\label{eq:alpha:beta:normal}
\vert\alpha_{\BRi,\BSi}\vert^2 - \vert\beta_{\BRi,\BSi}\vert^2 = 1.
\end{equation}
Then, the annihilation and creation operators defined by $\BRi_a$ can be written
in terms of the equivalent operators defined by $\BSi_a$, namely,
\begin{equation}
\begin{split}
\AO_\BRi &= \alpha^*_{\BRi,\BSi}\AO_\BSi + \beta^*_{\BRi,\BSi}\AO^{\dagger}_\BSi, \\
\AO_{\BRi}^{\dagger} &= \alpha_{\BRi,\BSi}\AO^{\dagger}_\BSi + \beta_{\BRi,\BSi}\AO_\BSi.
\end{split}
\end{equation}
We note that, if $\beta_{\BRi,\BSi}$ vanishes (or equivalently $\Comm{\AO_{\BRi,\HFi_1}}{\AO_{\BSi,\HFi_2}} = 0$), both sets define the same vacuum. In this case it is easy to see that $\BRi_a = e^{\ci\theta}\BSi_a$ for an arbitrary phase $\theta$. 

We express the vacuum defined by $\AO_{\BRi,\HFi}$ as $\ket{0_\BRi}$, such that $\AO_{\BRi,\HFi}\ket{0_\BRi} = 0$. Then, we define the mode number density operator
\begin{equation}\label{Nk}
N_{\BRi,\HFi} \equiv \AO^{\dagger}_{\BRi,\HFi}\AO_{\BRi,\HFi} = \BRi^{a}\BRi^{b*}\fsmvo_{\HFi ,a}(t_0)\fsmvo_{\HFi ,b}(t_0),
\end{equation}
and the total number operator
\begin{equation}
N_\BRi \equiv \modesum\;N_{\BRi,\HFi}
\end{equation}
such that its expected value at $\ket{0_\BSi}$ reads
\begin{equation}\label{eq:unit:evol}
\braketOP{0_\BSi}{N_\BRi}{0_\BSi} = \modesum\modedelta{\HFi}{\HFi}\left\vert\beta_{\BRi,\BSi}\right\vert^2 = \modesum\left\vert\beta_{\BRi,\BSi}\right\vert^2.
\end{equation}
We are assuming a compact hypersurface $\ST$, consequently the eigenfunctions are countable and $\modedelta{\HFi}{\HFi} = 1$.\footnote{For example, for a flat spatial hypersurface we can always choose a 3-torus, $\ST = \mathbb{T}^3$ with volume $L^3$, and take the $L\to\infty$ when necessary.} Finally, there is a simple relation between the two point function and the real part of the projector formed by $\BRi_a$, i.e., $\CS_{ab}$ defined in Eqs.~\eqref{defprp} and~\eqref{eq:def:CS}, namely
\begin{equation}\label{twop}
\braketOP{0_\BRi}{\hat\chi_a(x)\hat\chi_b(y)}{0_\BRi} = \modesum\HF{\HFi}(x)\HF{\HFi}(y)\frac{(\CS_{ab}+\SM_{ab})}{2},
\end{equation}
where $x,\,y \in \ST_{t_0}$.

As time evolves, the chosen basis will necessarily changes. Since the solutions of the equations of motion are unique given the initial conditions, the basis is determined for the whole time interval. Moreover, the creation and annihilation operators are defined through the conserved product Eqs.~\eqref{creation:operator} and~\eqref{eq:def:AO:fs}. Therefore, since the field operators satisfy the same equations of motion, the following product is conserved,
\begin{equation}
\label{annihilation-time}
\AO_{\HFi,\BR}(t) \equiv \BR^{a*}(t)\fsmvo_{\HFi ,a}(t) = \AO_{\HFi,\BRi}.
\end{equation}
The expression above shows that this particular combination of Heisenberg operators and time-dependent functions is the same at all time slices. Moreover, we can use this expression to write the field operator at any time $t$ as
\begin{equation}\label{at}
	\fsmvo_a(t) = \BR_a(t)\AO_\BRi + \BR^*_a(t)\AO^\dagger_\BRi.
\end{equation}
Naturally, due to Eq.~\eqref{annihilation-time}, the state annihilated by $\AO_{\HFi,\BR}(t)$ is the same at any time $t$.

Now, let us assume that we have a well defined physical prescription to determine the vacuum state at each time $t$, i.e., for each time slice we define a normalized phase vector $V_a(t)$ such that the annihilation operator
\begin{equation}
\AO_{V}(t) \equiv	V^{a*}(t)\fsmvo_{\HFi ,a}(t)
\end{equation}
defines the physical vacuum state $\AO_{V}(t)\ket{0_V(t)} = 0$. Notice that $V_a(t)$ is a time-dependent phase vector, but it does not necessarily satisfy the equations of motion. Consequently neither $\AO_{V}(t)$ nor $\ket{0_V(t)}$ will be inevitably constant. We denote this prescription as a Vacuum Determination Criterion (VDC). 

Using a VDC, one can construct the vacuum state (and consequently the whole Fock space) at each time-slice as $\ket{0_V(t)}$. Suppose that at $t_0$ we assert that the quantum field is found at the vacuum $\ket{0_V(t_0)}=\ket{0_v}$. This is implemented by choosing $\BRi_a = v_a\equiv V_a(t_0)$ as the initial conditions for the phase vector. However, since $V_a(t)$ is not a solution of the field equations, in general $V_a(t)$ and $\BR_a(t)$ (which is the dynamically time evolved vacuum from $\BRi_a = v_a$) may differ at $t\neq t_0$. Consequently, if we measure the number density operator 
\begin{equation}
N_{\BR,\HFi}(t) \equiv \AO^{\dagger}_{\BR,\HFi}(t)\AO_{\BR,\HFi}(t),
\end{equation}
on the vacuum state at $t$, i.e., $\ket{0_V(t)}$, we get
\begin{align}
\braketOP{0_V(t)}{N_{\BR,\HFi}(t)}{0_V(t)}	&= \left\vert\beta_{\BR,V}(t)\right\vert^2, \\ \beta_{\BR,V}(t) &= V_a(t)\SM^{ab}\BR_b(t).
\end{align}
Hence, the expectation value above is zero at $t_0$, while for $t \neq t_0$ it depends on how much $\BR_a(t)$ differs from $V_a(t)$. Note that if they differ only by a phase, then $\beta_{\BR,V}(t)=0$.

In practice, a vacuum prescription provides a set of four real numbers for each mode $\HFi$, i.e., the two complex components of $v_a$ in the complex phase space $\PSV$. However, two vacuum prescriptions that differ only by a phase specify the same vacuum, i.e., they are equivalent. Physically we are only interested in the class of equivalence $[\BR_a]$ of phase vectors related by the equivalence relation $\BR_a\sim V_a$ iff $\BR_a = e^{i\theta(t)}V_a$, for any real function $\theta(t)$. Moreover, if the normalization condition $\BR^{a*}\BR_{a} = 1$ is satisfied by a phase vector, then it is satisfied by any other vector in $[\BR_a]$. Thus, a VDC determines classes of equivalence of normalized phase vectors belonging to the reduced phase space $\PSVr$ of phase space vectors modulo a phase.\footnote{The terminology ``reduced phase space'' we are using here has nothing to do with the reduced phase space defined in the context of constrained Hamiltonian systems.} In addition, once a VDC is chosen, we need to compute the time evolution using Eq.~\eqref{eqmovY}. These equations preserve the normalization condition, and it is easy to see that two different initial conditions taken from the same equivalence class $[\BR_a]$ at $t_0$ evolve to the same class of equivalence at $t$, i.e, if $\BRi_a \sim \BSi_a$ at $t_0$, then $\BR_a(t) \sim \BS_a(t)$. In this sense, the vacuum dynamics also happen in $\PSVr$, whose structure will be explored in the following subsection.

\subsection{Representations of the reduced phase space}
\label{subsec:repr}

It is useful to construct different representations of the reduced phase space. In App.~\ref{app:Mv}, we prove that there is a one-to-one mapping between $\PSVr$ and the space of linear complex structures $\PSVM$ represented by real matrices $M_a{}^b$. Furthermore, in App.~\ref{app:alphabeta} we show that $\PSVM$  is a two-dimensional manifold, and we obtain the $(\alpha, \gamma)$ parametrization mapping the whole manifold into $\mathbb{R}^2$. The dynamics induced in $\PSVM$ by Eq.~\eqref{eqmovY} can be readily obtained using the mapping~\eqref{eq:def:CS} from the field variables to $M_a{}^b$. It is more convenient to write the equations using the Clifford algebra representation as defined in App.~\ref{app:matrix} and~\ref{app:ct}. Since the components $\gH^{ab}$ are real, the tensor $N_a{}^b \equiv -\ci\gH_a{}^b$ is also real (see Eq.~\eqref{defN} for its specific form). Then, the equation of motion for $M$ is
\begin{equation}\label{eqmovM}
\dot{M} = 2N\wedge M.
\end{equation}
When expressed in terms of the parametrization given in Eq.~\eqref{eqM}, it reads
\begin{equation}\label{eqmovAG}
\begin{split}
\dot{\alpha} &=-2\nu\sh(\gamma-\xi),\\
\dot{\gamma} &=+2\nu\ch(\gamma-\xi)\tah(\alpha) - 2h,
\end{split}
\end{equation}
where $\nu,\;\xi$ and $h$ are given functions of time coming from the Hamiltonian and composing $N$, while $\alpha$ and $\gamma$ are the dynamical variables belonging to $M$. In this representation, we now have two non-linear ordinary differential equations instead of four linear but constrained dynamics. 

Using the same mapping between $\PSVr$ and $\PSVM$, a VDC in this representation will be just a time dependent linear structure $V(t) \in \PSVM$ (with associated phase vector $v_a$), which does not necessarily satisfies the equation of motion \eqref{eqmovM}. Defining the vacuum at $t_0$ using $V(t_0)$, its evolution in time will be given by Eq.~\eqref{eqmovM}, with the initial condition $M(t_0) = V(t_0)$. This procedure specifies a unique $M(t)$ and, consequently, the associated phase vector $\BR_a(t)$. Usually, $M(t)$ is different from $V(t)$. In this case, the amount of particles created at $t$ with respect to this VDC is given by the Bogoliubov coefficients between $\BR_a(t)$ and $V_a(t)$, which in the $\PSVM$ representation ($R_a(t)\rightarrow M(t)$ and $V_a(t) \rightarrow V(t)$) reads
\begin{align}
\nonumber\left\vert\alpha_{\BR,v}(t)\right\vert^2  &= V^*_a(t)\SM^{ab}\BR_b(t)\BR^*_c(t)\SM^{cd}V_d(t),\\
\nonumber &= +\frac{1}{4} \mathrm{Tr}\left[{I} - M(t)V(t)\right], \\
&= \frac{1-M(t)\cdot V(t)}{2},\\
\nonumber\left\vert\beta_{\BR,v}(t)\right\vert^2 &= V_a(t)\SM^{ab}\BR_b(t)\BR^*_c(t)\SM^{cd}V^*_d(t), \\
\nonumber &=  -\frac{1}{4} \mathrm{Tr}\left[I + M(t) V(t)\right],\\ 
\label{betaK} &=-\frac{1+M(t)\cdot V(t)}{2}.
\end{align}
For clarity, we wrote above the expressions for the Bogoliubov coefficients in all three representations: phase vector, matrix, and Clifford algebra, respectively. As already noticed in the preceding sub-section, when $V(t)$ does not satisfy the equations of motion, the Bogoliuov coefficients are not constant in time. They measure the particle creation at $t$ given the vacuum set at $t_0$ using the VDC $V(t_0)$.

It is worth noting that, an element of $\PSVM$ is just a ``time''-like vector $\vm$ satisfying $\vm^2=\vm\cdot \vm =-1$, and positive ``time'' component $-\vg_0\cdot\vm>0$. This means that $\PSVM$ has a natural mapping to the hyperbolic space $\mathbb{H}^2$ through the hyperboloid model as discussed in App.~\ref{app:hyspace}. Thus, given two ``time''-like unitary vectors $\vm$ and $\vv$, the $\beta$ coefficient associated with the transformation between them (Eq.~\eqref{betaK}) is just
\begin{equation}
	\vert\beta_{\vm,\vv}|^2 = -\frac{1}{2}\left(1+\vm\cdot \vv\right) = \left(\frac{\vm-\vv}{2}\right)^2,
\end{equation}
where we are using $\beta_{\vm,\vv}$ as a synonym to $\beta_{\BR,v}$. The equation above shows that $\beta_{\vm,\vv}$ is just one-half of the norm of the vector difference between $\vm$ and $\vv$. Nevertheless, this is not a metric but a semi-metric: it does not satisfy the triangle inequality (actually, it satisfies the reverse triangle inequality). The genuine distance in $\mathbb{H}^2$ is presented in Eq.~\eqref{def:eta}: given two ``time-like'' points $\vm$ and $\vq$ in $\mathbb{H}^2$, the hyperbolic distance between these two points, which is a proper metric, reads  $d(\vm,\vv) \equiv \acosh\left(-\vm\cdot \vv\right)$. Therefore, the Bogoliubov coefficients have a very simple relation with the distance $d(\vm,\vq)$:
\begin{equation}
\label{beta-distance}
\vert \alpha_{\vm,\vv} \vert = \ch\left[\frac{d(\vm,\vv)}{2}\right],\; 
\vert \beta_{\vm,\vv} \vert = \sinh\left[\frac{d(\vm,\vv)}{2}\right].
\end{equation}
For convenience, we also define the distance between the unit ``time'' vector
$\vg_0$ and an arbitrary multi-vector $\vm$ as
\begin{equation}
d_\vm \equiv \acosh\left(-\vm\cdot \vg_0\right).
\end{equation}

\section{Vacuum determination criteria}
\label{VDC}

The sequence of definitions above shows that a choice of phase vectors specifies a set of creation and annihilation operators, which in turn defines a vacuum state and all other particle states using their commutation relations. Given a VDC, one can compute the particle creation as a function of this choice. However, it does not seem possible to provide a VDC in a single time slice, as, in principle, any phase vector provides a well-defined representation. Therefore, a physically well-defined vacuum state should be related to the field evolution. In this section, exploring the dynamics of the field, we will present the strategy to select a particular phase space basis and its associated vacuum state in the simple and well-known case of a time-independent Hamiltonian. Furthermore, using the representations of the reduced phase space introduced in Sec.~\ref{subsec:repr}, we present an equivalent new viewpoint on this strategy, which will be used to extend it to the time-dependent case in the following section.

\subsection{Time-translation symmetry in the dynamics and stability in the phase vector representation}

We previously showed that two phase vectors differing by a phase define the same vacuum state. Thus, if the time evolution of a particular phase vector is such that only its phase change with time, then this choice of vector would define the same vacuum, independently of the chosen time slice. That should be the case when the physical laws governing the dynamics of the quantum field are invariant under time translations, as it would impose that its natural quantum vacuum state should be time-independent. Hence, a natural VDC emerges in this situation: the phase vector determining the vacuum state is the same at all times, apart from a phase, 
\begin{equation}
\label{vacuum-t-ind}
V_a(t)=\exp[-i\varphi_1(t)] v_a,
\end{equation} 
where $v_a\equiv V_a(t_0)$, and $\varphi_1(t)$ is an arbitrary function of time, with $\varphi_1(t_0)=0$ for convenience. 

However, the vector $v_a$ is still completely arbitrary, and we have to fix it, modulo a phase. One second criterion could be to demand stability of the vacuum state, i.e., that the dynamically time evolved $v_a$, $\BR_a(t)$, coming from Eq.~\eqref{eq:motion:BF} with $\BR_a(t_0)=v_a$,
\begin{equation}
\label{HamN}
\dot\BR_a(t)  = -\ci\SM_{ac}\gH^{cb}(t)\BR_b = \sigma(t) U_a{}^b(t) \BR_b,
\end{equation}
should also differs from $V_a(t)$ at most by a phase, 
\begin{equation}
\label{vacuum-ind-evol}
\BR_a(t)=\exp[-i\varphi_2(t)] V_a(t)  ,
\end{equation} with $\varphi_2(t)$ also being an arbitrary function of time.\footnote{Note that the stability criterion is connected with the requirement that a vacuum state should be the minimum energy state. Because, contrary to the excited states, a minimum energy state is stable when interactions with external fields are negligible.} In Eq.~\eqref{HamN} we define the real Hamiltonian tensor as in Eq.~\eqref{defN}, i.e.,
\begin{equation}\label{rht}
N_a{}^b(t) \equiv -\ci\SM_{ac}\gH^{cb}(t), \quad U_a{}^b \equiv \frac{N_a{}^b}{\sigma(t)}.
\end{equation}
Equations \eqref{vacuum-t-ind} and \eqref{vacuum-ind-evol} imply that
\begin{equation}\label{phaseonly}
\ci\dot{\BR}_a = \omega(t) \BR_a,
\end{equation}
for an arbitrary real function $\omega(t) = \dot{\varphi}_1(t) + \dot{\varphi}_2(t)$.
Equating Eqs.~\eqref{phaseonly} and \eqref{HamN}, we get that $\BR_a$ must be an eigenvector of $N_a{}^b(t)$, i.e.,
\begin{equation}\label{eigen}
U_a{}^b(t) \BR_b = -\ci\BR_a,
\end{equation}
where we have set $\omega(t) = \sigma(t)$.  

In App.~\ref{app:HT} we show that $N_a{}^b(t)$ have two normalizable eigenvectors with purely imaginary eigenvalues iff $\gH^{ab}(t)$ is positive definite, which we will assume to be true here. However, since Eq.~\eqref{eigen} defines $\BR_a(t)$ within a phase, $\BR_a(t)$ will satisfy both Eqs.~\eqref{HamN} and~\eqref{eigen} iff $\dot{U}_a{}^b=0$, as expected.\footnote{That can be easily checked by differentiating Eq.~\eqref{eigen} with respect to time.} Hence, we obtain the well-known result that a globally stable vacuum state can be reached only when the background geometry, where the field evolves, has a time-like Killing vector field, and no time-dependent interactions are present. That is the case of free fields in flat and de Sitter spacetimes, yielding the so-called Minkowski and Bunch-Davies vacua \cite{Bunch1977,Birrell1982,Allen1985}, respectively. Therefore, in this particular situation, one can establish as a vacuum determination criterion (VDC) the (constant, globally defined) eigenvector of $N_a{}^b(t)$ at any time $t$, $\BS_a$, which selects $v_a$ in Eq.~\eqref{vacuum-t-ind} as 
 \begin{equation}\label{VDC1}
v_a = \BS_a, \quad U_a{}^b\BS_b=-\ci \BS_a.	
 \end{equation}
Applying this VDC at $t_0$ and evolving to $t$ using the dynamical equations \eqref{HamN} yields
\begin{equation}
\BR_a(\tau) = e^{-\ci(\tau-\tau_0)} v_a = e^{-\ci(\tau-\tau_0)} \BS_a,
\end{equation}
where  we define a new time variable $\tau = \int \sigma(t) \ddt$. From the VDC Eq.~\eqref{vacuum-t-ind}, one gets
\begin{equation}
V_a(\tau) = e^{-\ci\varphi_1(\tau)} v_a(\tau) = e^{-\ci\varphi_1(\tau)} Q_a.
\end{equation}
Accordingly, the particle creation number density is $\beta_{\BR,V}(t) = 0$, since $\BR_a(\tau)$ and $V_a(\tau)$ differ by a phase, and the vacuum states defined by both vectors are equivalent. 

We can trivially go back to the original time $t$, obtaining
\begin{equation}
\label{vacuum1}
\BR_a(t) = \exp\left(-\ci\int_{t_0}^{t} \sigma(t')\ddt'\right)\BS_a .
\end{equation}
For a Hamiltonian tensor parametrized by
\begin{equation}\label{Horig0}
\gH^{ab} \doteq \left( \begin{array}{cc} m\nu^2 & h \\ h & \frac{1}{m} \end{array} \right),
\end{equation}
where $m$, $\nu$ and $h$ are scalar functions, the eigenvector (defined within an arbitrary phase) $\BS_a$ of $U_{a}{}^b$ reads (see Eq.~\eqref{eivenVN})
\begin{equation*}
\BS_a \doteq \left(-\frac{e^{-\ci\varphi_h}}{\sqrt{2m\nu(1-h^2/\nu^2)}},\;e^{\ci\varphi_h}\sqrt{\frac{m\nu}{2(1-h^2/\nu^2)}}\right),
\end{equation*}
with $\varphi_h \equiv \tan^{-1}\sqrt{(\nu-h)/(\nu+h)}$, and $\sigma = (\nu^2-h^2)^{1/2}$. For Hamiltonians with $h=0 \Rightarrow \sigma =\nu$, the phase vector reduces to a more familiar form, namely
\begin{equation}
\BS_a \doteq -e^{-\ci\pi/4}\left(\frac{1}{\sqrt{2m\nu}},\;-\ci\sqrt{\frac{m\nu}{2}}\right).
\end{equation}
For example, in the case of a free mass-less scalar field in a Minkowski space-time, $m=1$ and $\nu = \EV{\HFi} = \vert\HFi\vert$, yielding
$$\BR_a(t) = e^{-\ci\vert\HFi\vert(t-t_0)}\BS_a .$$
Note that the VDC oscillates in time in $\PSV$, but it is a fixed point in $\PSVr$.

It is important to emphasize here that the time-translation symmetry implying that the quantum vacuum state should be time independent (modulo a phase) was not only insufficient to fix the vacuum, but it also does not lead to a stable vacuum state in general. Indeed, any other choice of a constant $V_a(\tau)$ satisfying the time-symmetry criterion not proportional to $\BS_a$ is not stable. For instance, take another constant $V_a(\tau)$ written in terms of the basis formed by $\BS_a$ and $\BS_a^*$, as,
\begin{equation}
V_a(\tau) = c_1 \BS_a-c_2 \BS_a^*,
\end{equation}
where $c_1$ and $c_2$ are arbitrary complex constants that must be compatible with the normalization of $V_a(\tau)$, i.e., 
$$\vert c_1\vert^2 - \vert c_2\vert^2 = 1.$$
Applying the time evolution we get
\begin{equation}
\BR_a(\tau) = c_1 e^{-\ci\Delta\tau} \BS_a-c_2 e^{+\ci\Delta\tau} \BS_a^*,
\end{equation}
where $\Delta\tau \equiv \tau - \tau_0$. Expressing this solution in terms of the VDC at $\tau$ we get
\begin{equation}
\begin{split}
\BR_a(\tau) &= \left[\cos\Delta\tau-\ci\left(\vert \ c_1 \vert^2+\vert c_2\vert^2\right)\sin\Delta\tau\right]v_a \\
&-2\ci c_1 c_2 \sin\Delta\tau \;v_a^*,
\end{split}
\end{equation}
and $\vert\beta_{\BR,V}(t)\vert = \vert 2 c_1 c_2 \sin\Delta\tau\vert$.
Hence, there is a time oscillating particle production, and a static VDC does not generically lead to a stable vacuum, unless $c_2=0$. That brings us back to the well-known vacuum state of the time-independent dynamical system presented above. 

\subsection{The matrix and Clifford algebra representations: a new viewpoint}
\label{constcase}

It is useful to reexamine the example discussed above in the matrix and Clifford algebra representations. In this case the Hamiltonian tensor is such that $U$ (see Eq.~\eqref{rht}) is constant. Thus, the respective equation of motion reads
\begin{equation}\label{eqmo0}
	\dot{M} = 2\sigma\; U\wedge M.
\end{equation}
The solution can be readily obtained using the exponential map~\eqref{exp:v}, that is, 
\begin{equation}\label{solUconst}
M = e^{\Delta\tau U} M_0 e^{-\Delta\tau U},
\end{equation}
where $M_0$ is the initial condition. If the Hamiltonian is positive definite then $U^2=-1$, hence positive definite Hamiltonians lead to a ``time''-like multi-vector $U$. Rewriting $M_0$ in terms of $U$ as described in Eq.~\eqref{decompvp} leads to
\begin{equation}
\label{M-static}
\begin{split}
M &=  \cosh d(M_0,U)\,U \\
&+\cos\left({2\Delta\tau}\right){M_0\perp U} -\sin\left({2\Delta\tau}\right){M_0\wedge U}.
\end{split}
\end{equation}
This last expression shows that all solutions in the reduced phase space stay at fixed distances from $U$, i.e., $M \cdot U = M_0\cdot U$ is constant, and rotate around $U$ with period $\pi$, which is twice as fast as the oscillations in phase space. The point $U$ is stable concerning the solutions given that their distances to this point do not change in time. In particular,  the choice $M_0=U$ implies $M=U$, and this solution does not oscillate at all. In other words, when $U$ is constant, $M=U$ is a solution of Eq.~\eqref{eqmo0}, being a fixed point of the reduced phase space and the dynamics. 
   
It is also informative to examine this problem from the point-of-view of Eq.~\eqref{eqmovAG}. For constant $\xi$ and $\;h/\nu$ the point 
$$\gamma=\xi,\qquad \alpha=\tanh^{-1}\left(\frac{h}{\nu}\right) $$
is a fixed point of this autonomous system. These are the same values we get by setting $M_0 = U$. Hence, establishing the VDC as $V(t)=U$, which is tantamount to make the choice of $V_a(t) = \BS_a$ (modulo a phase) as an eigenvector of $U_a{}^b$ (see Eq.~\eqref{VDC1}), yields a globally stable vacuum because it is a fixed point of Eq.~\eqref{eqmovAG}, or a solution of Eq.~\eqref{eqmo0}, $M=U=V(t)$. Therefore, using a different representation, we arrive at the same final result as before in the case $U$ is constant. However, this new form of presenting the vacuum state sheds light into a new important property of this prescription: the vacuum choice is the unique point in the reduced phase space where all solutions stay at constant distance from it.

In summarizing, in the case of a time-independent Hamiltonian, we could obtain a vacuum state which is a fixed point in the reduced phase space and the dynamics. This single fixed point leads to the vacuum prescription $V(t)=U$, which results in no particle creation. Any other constant VDC would result in particle creation with a $\beta_{\vm,\vv}$ given by $\sh[d(\vm,\vv)]$. Of course, the choice that minimizes the particle creation is exactly $\vv = U$. A remarkable property of this choice is that all solutions are stable orbits around $U$, with a fixed distance from them.
In the general case, however, the Hamiltonian is time-dependent. Therefore, the system no longer has a fixed point. Nonetheless, we can still define a notion of stability motivated by the time-independent case and the new property of its vacuum described above: given a set of solutions of the dynamical system, one should seek for a stability curve $V(t)$ whose distance to any arbitrary solution in the reduced phase space varies very slowly. Moreover, solutions draw curves with smaller lengths as they get closer to the trajectory. Consequently, if we choose such $V(t)$ as our VDC, then the solution $M(t)$ with initial condition $V(t_0)$ remains close to $V(t)$, as the distance $d(M(t),V(t))$ varies slowly and $d(M(t_0),V(t_0))=0$, implying that $d(M(t),V(t))\ll1$. Therefore such stability curve defines a physical vacuum, with negligible particle production, $\vert \beta_{\vm,\vv} \vert = \sinh\left[\frac{d(M(t),V(t))}{2}\right]\ll1$. We present concrete examples of these stability curves in the next sections. In the adiabatic case, the stability curve reduces to the well-known adiabatic vacuum, but we also show that in the non-adiabatic regime this stability curve can still be found and provide a physically motivated VDC.


\section{The case of a general time-dependent Hamiltonian}
\label{sec:gcase}

For a generic Hamiltonian dependent on time, the vector $\vu$ moves in the reduced phase space, and it is not a solution of Eq.~\eqref{eqmovM}. Thus, even setting $\vm(t_0) = \vu(t_0)$, $\vu(t)$ moves differently on $\PSVr$ than $\vm(t)$ ($\vm(t)$ is a solution of Eq.~\eqref{eqmovM}, while $\vu(t)$ is not). Their distances changes in time, and particle creation generally takes place. However, as described at the end of the previous section, even in the time-dependent case there are situations where stability curves can be found and used as a VDC. 
In this section, we present two very important physically motivated examples where this procedure can be successfully implemented. 

\subsection{The adiabatic case}
\label{subsec:nu}

One first possibility is that $\vu$ moves slowly compared to the frequency of the circles around $\vu$ drawn by $\vm$. These circles are dragged by the movement of $\vu$, thus keeping the solution around $\vu$ for a long period. Exploring this feature of such type of dynamics is the essential tool to find stability curves in the reduced phase space and construct the so-called adiabatic vacuum.

To find stability curves, one first has to study the solution space of the dynamical system. In the previous sections, it was possible to find them because the analytical solution is attainable when the multi-vector $U$ is constant. In this case we can write the propagator explicitly as in Eq.~\eqref{solUconst},
\begin{equation}\label{propUconst}
P(t,t_0) = e^{\Delta\tau U}, \quad P^\dagger(t,t_0) = e^{-\Delta\tau U}.	
\end{equation}
When $\vu$ depends on time, $\int\dd t \, \vn$ and $\vn$ do not commute in general. For this reason, there is no such simple solution for the propagator equation,
\begin{equation}
\dot {P}(t) =N(t)P(t).
\end{equation}

This problem can be circumvented by making successive canonical transformations on the system such that $\vu$ is transformed as close as possible to an almost constant multi-vector. Then, in the frame where $U$ is well approximated by a constant multi-vector, we can compute the solutions similarly to Eq.~\eqref{solUconst}.

As described in App.~\ref{app:ct}, after a canonical transformation $C$, the transformed dynamical vector $\vm^C$ and the new Hamiltonian $\vn^C$ read
\begin{align}
\label{TransformedVectors}
\vm^C&\equiv C\vm C^\dagger, \nonumber &\\ 
\vn^C &\equiv \sigma C\vu C^\dagger+\sigma \delta \vu^C, & \delta \vu^C &\equiv \frac{\dot{C}C^\dagger}{\sigma},
\end{align}
while Eq.~\eqref{eqmovM} retains its form,
\begin{equation}
\dot{\vm}^C = 2\vn^C\wedge \vm^C,
\end{equation}
where we used that $CC^\dagger = 1$ and $\dot{C}C^\dagger + C\dot{C}^\dagger = 0$. Note that $\vn^C$ is not normalized, not only because of $\sigma$, but also due to the presence of $\delta \vu^C$, which is the well-known additional term that a time-dependent canonical transformation introduces in the transformed Hamiltonian.

Assuming that the Hamiltonian is positive-definite, implying that the vector $\vu$ is ``time''-like, the natural choice to transform $\vu$ is to take it as close as possible to $\vg_0$,\footnote{In principle, we could transform $U$ into any unity ``time''-like multi-vector. We choose $\vg_0$ for simplicity.} that is, we go to a frame where $\vu$ has negligible ``spatial'' components. Equivalently, we are transforming $\mathbb{H}^2$ such that $\vu$ points as near as possible to the origin of the coordinate system (the point $\vg_0$). It is easy to check that the following canonical transformation takes $\vu^C\equiv C_0\vu C_0^\dagger = \vg_0$,
\begin{equation}\label{CTug0}
\begin{split}
C_0 &= \exp\left(\frac{d_{\vu}}{2}\vu_{s}\right), \qquad\qquad\;\qquad \vu_{s} \equiv \widebar{\vu\wedge \vg_0},\\
C_0 &=  \cosh\left(\frac{d_{\vu}}{2}\right) + \sinh\left(\frac{d_{\vu}}{2}\right)\vu_{s}.
\end{split}
\end{equation}
Note, however, that the additional term $\delta \vu^C$ is necessarily present in the new Hamiltonian when $\dot{\vu}\neq 0$. This term can be computed explicitly as
\begin{equation}\label{du1}
\delta \vu^C = \frac{\dot{d}_{\vu}}{2\sigma}\vu_{s}+\sinh d_{\vu}\frac{\dot{\vu}_{s}}{2\sigma}-\sinh^2\left(\frac{d_{\vu}}{2}\right)\frac{\dot{\vu}_{s}\vu_{s}}{\sigma}.
\end{equation}

In the case where $\vu$ changes adiabatically (e.g., the time derivative of $\vu$ is smaller than $\sigma$, or the time scale in which $\vu$ changes is much larger than the time scale arising from $\sigma$), all components of $\vu$ satisfy
\begin{equation}\label{adiabcond}
\frac{1}{\sigma}\frac{\dd}{\dd t} \vu^A \ll 1,\quad \left(\frac{1}{\sigma}\frac{\dd}{\dd t}\right)^{N+1} \vu^A \ll \left(\frac{1}{\sigma}\frac{\dd}{\dd t}\right)^N \vu^A,
\end{equation}
for all positive integers $N$ up to $N_\mathrm{max}$. Then $\delta \vu^C$ can be seen as a first order adiabatic correction to $\vu^C$. We will denote it as $\delta \vu^{(1)}$, as well as any other object resulting from this first canonical transformation. In this way, the new Hamiltonian vector reads
\begin{equation}
\vn^{(1)} \equiv \sigma (\vg_0 + \delta \vu^{(1)}),
\end{equation}
while the new equation of motion is
\begin{equation}\label{eqm1a}
\dot{\vm}^{(1)} = 2\vn^{(1)} \wedge \vm^{(1)}.
\end{equation}
In terms of the normalized Hamiltonian
$$\vu^{(1)} \equiv \frac{\vn^{(1)}}{\sqrt{-\vn^{(1)2}}} \approx \vg_0 + \delta \vu^{(1)}\!\!\perp\!\vg_0,$$
and the new frequency
$$\sigma_1 \equiv \sqrt{-\vn^{(1)2}} \approx \sigma\left(1-\delta\vu^{(1)}\cdot\vg_0\right),$$
the equations of
motion~\eqref{eqm1a} are now
\begin{equation}
\dot{\vm}^{(1)} = 2\sigma_1 \vu^{(1)} \wedge \vm^{(1)}.
\end{equation}
It is worth pointing out here that this normalization is possible if the
adiabatic conditions~\eqref{adiabcond} are satisfied. In other words, since $\delta\vu^{(1)}$ has modulus much smaller than one, the ``time''-like property of $U$ is unchanged. We refer to this frame as the first order adiabatic frame.

The second order adiabatic frame can be obtained in a similar way: we find the canonical transformation that takes $\vu^{(1)} \to \vg_0$, and apply it to the equations of motion above. In this case, $\delta \vu^{(2)}$ is composed by time derivatives of $\vu^{(1)}$,  which are necessarily second order. The new $\vn^{(2)}$ is then normalized, yielding the new equation
\begin{equation}
\dot{\vm}^{(2)} = 2\sigma_2 \vu^{(2)} \wedge \vm^{(2)}.
\end{equation}
One can then proceed in the same way up to $N_\mathrm{max}$. Too get back to the original frame one only needs to perform the respective inverse canonical transformations. In App.~\ref{app:adiab} we develop a recursive method to compute all canonical transformations up to $N_\mathrm{max}$.

Up to this point we introduced the adiabatic frames without any approximation. To study the space of solution in a frame ${}^{(n)}$ we can compute the approximate propagator, that is,
\begin{equation}
\label{prop-ad}
\dot {P}^{(n)}(t) =N^{(n)}(t)P^{(n)}(t).
\end{equation}
In this frame the Hamiltonian multi-vector reads 
\begin{equation}
\label{NU-n}
	\vn^{(n)}(t)  = \sigma_{n-1}(t)\left(\vg_{0} - F_n(t)\vg_n\right),
\end{equation}
where $\vg_n$ is either $\vg_1$ or $\vg_2$ defined in Eq.~\eqref{g0-g1-g2}, depending on the order $n$.
The functions $F_n(t)$ and $\sigma_{n-1}(t)$ are given in Eq.~\eqref{boostparams} of Appendix C,
\begin{equation}
	\begin{split}
		F_n &= (-1)^n\frac{\dot{\xi}_{n-1}}{2\sigma_{n-1}}, \qquad \xi_{n} = \tanh^{-1}\left(F_n\right),\\
		\sigma_{n} &= \sigma_{n-1}\sqrt{1-F_{n}^2},
	\end{split}
\end{equation}
where the initial functions of these recurrence relations are
\begin{equation}
	\sigma_{0} \equiv \sigma, \qquad \xi_{0} \equiv \xi = \ln(m\sigma).
\end{equation}
Note that, because of the adiabatic evolution, $|F_n|\ll 1$ and $U^{(n)}\approx\vg_{0}$. 

In App.~\ref{app:adiab}, we show how to obtain a simple approximate propagator up to order $n$ coming from Eqs.~\eqref{NU-n} and \eqref{prop-ad}). It reads
\begin{equation}\label{proporder0}
P^{(n)}(\tau,\tau_0) \approx e^{-p_r^{(n)}(\tau)}e^{\Delta\tau\vg_0},
\end{equation}
where the time-dependent ``space-vector'' $p_r^{(n)}(\tau)$ reads.
\begin{equation}
p_r^{(n)}(\tau) \approx \frac{F_n(\tau)}{2}\vg_0\wedge\vg_{n}.
\end{equation}
The expression above shows that the time evolution (up to order $n$) is a rotation about the origin $\vg_0$ followed by a boost in the $\vg_0\wedge\vg_{n}$ direction. 

We can now apply the propagator~\eqref{proporder0} to an arbitrary initial condition $\vm_0^{(n)}$. First, we rewrite the multi-vector $\vm_0^{(n)}$ as 
\begin{equation}\label{inicond}
\vm_0^{(n)} = \cosh r\,\vg_0 + \sinh r\, \bar{m},
\end{equation}
where $\bar{m}$ is a purely ``spatial'' unitary vector, i.e., $\vg_0\cdot \bar{m} = 0$ and $\bar{m}^2=1$ and $r$ in an arbitrary initial distance from $\vg_0$. The first factor of the propagator is simply the rotation about the origin,
\begin{equation}
	e^{\Delta\tau\vg_0}\vm_0^{(n)}e^{-\Delta\tau\vg_0} = \cosh r\,\vg_0 + \sinh r \,\bar{m}_r,
\end{equation}
where $\bar{m}_r$ is a rotating unitary ``spatial'' vector,
\begin{align}
\label{m-rotated}
\bar{m}_r &= e^{\Delta\tau\vg_0} \,\bar{m}\,e^{-\Delta\tau\vg_0}\nonumber \\ 
&= \cos(2\Delta\tau) \bar{m}+ \sin(2\Delta\tau) \vg_0\wedge\bar{m}.
\end{align}
Before applying the second factor of the propagator, it is convenient to decompose the vector $m_r$ in normalized components parallel and perpendicular to the normalized $\bar{p}_r^{(n)}$ instead of $\bar{m}$ (as in~\eqref{normalized}), that is,
\begin{equation}
\label{expansion}
\bar{m}_r = c_1(\tau) \bar{p}_r^{(n)}+ c_2(\tau) \vg_0\wedge \bar{p}_r^{(n)}.
\end{equation}
As $\bar{m}_r$ is normalized, $c_1^2(\tau) + c_2^2(\tau)=1$. Note that $c_1(\tau)=\bar{m}_r\cdot \bar{p}_r^{(n)}\equiv \cos(\theta(\tau))$, hence $c_2(\tau)= \sin(\theta(\tau))$.

Finally, we have
\begin{align}
\label{M-nonstatic}
\vm^{(n)} &= e^{-p_r^{(n)}(\tau)}e^{\Delta\tau\vg_0} \vm_0^{(n)}e^{-\Delta\tau\vg_0}e^{p_r^{(n)}(\tau)},\\
\nonumber &=\left(\cosh r\,\cosh d_p + \sinh r\,\sinh d_p\sin\theta\right)\vg_0\\
\nonumber &+\left(\cosh r\,\sinh d_p + \sinh r\,\cosh d_p\sin\theta\right)\vg_0\wedge\bar{p}_r^{(n)}\\
&+\sinh r\,\cos\theta\;\bar{p}_r^{(n)},
\end{align}
where
\begin{equation}
d_p(\tau)\equiv 2\left\vert p_r^{(n)}(\tau)\right\vert=|F_n(\tau)|.
\end{equation}
As the solutions evolve, the distance between $\vm$ and the origin, $d_M = \cosh^{-1}(-\vm\cdot\vg_0)$, varies within
\begin{equation}
\vert r - d_p\vert \leq d_M\leq \vert r + d_p\vert ,
\end{equation}
as $\theta$ varies from $-\pi/2$ to $\pi/2$.

First note that when $d_p(\tau)=|F_n(\tau)|=0$, and using Eqs.~\eqref{m-rotated} and \eqref{expansion}, we recover the time-independent case Eq.~\eqref{M-static} from Eq.~(\ref{M-nonstatic}) if $U$ is boosted to $\vg_0$ (which is trivial when $U$ is constant). In this case, we notice all solutions rotate at a constant distance $r$ from $U=\vg_0$. Therefore the curve $V^{(n)}(t) = \vg_0$ is a perfect static curve, yielding a globally stable vacuum with no particle creation as long as $r =d_{V^{(n)}(t_0)}=0$. That is our VDC in that situation.

In the adiabatic case, when $d_p(\tau)=|F_n(\tau)|\neq 0$, the solutions rotate around $\vg_0$ inside an annular disk with maximum width $2 d_p(\tau)$, and maximum radius $r + d_p(\tau)$. In this case, the curve $V^{(2)}(t) = \vg_0$ is not a perfectly static curve because even with $r=0$ solutions still rotate around $\vg_0$ at a distance $d_p(\tau)\neq 0$ from it. However, in the adiabatic approximation $d_p(\tau) \ll  1$, hence we can consider the curve $V^{(n)}(t) = \vg_0$ as a quasi-static curve: during the time interval in which the adiabatic approximation is valid, the solutions remain close to it. Therefore, we choose $\vg_0$ as our VDC, note however that this is a frame dependent choice, if the VDC is set at frame $(n)$ then $V^{(n)}=\vg_0$ but if computed at a different frame it will differ from $\vg_0$, for example for $0<n_0<n$ we have $$V^{(n_0)} = \vg_0 + \sum_{j=n_0+1}^{n}\bigO{j}.$$ Going to the original frame the last canonical transformation is not necessarily small since it is generated by $\xi$, thus, the VDC at the original frame has a potentially large boost in the $\vg_2$ direction. When necessary, we write $V_n^{(n_0)}$ to denote the VDC set at frame $n$ but computed at frame $n_0$, naturally $V_n^{(n)} = \vg_0$. These points are illustrated at Sec.~\ref{sec:adiab} where we present an example of the VDC in a cosmological setting.

Now, let $\vm^{(n)}(t)$ be the time evolved solution from $V_n^{(n)}$, the Bogoliubov coefficient for particle production reads
\begin{equation}
\label{beta-distance-Ad}
\begin{split}
\vert \beta_{\vm^{(n)},V^{(n)}}(t) \vert &= \vert \beta_{\vm,V}(t) \vert\\
&= \sinh\left[\frac{d_{\vm^{(n)}(t)}}{2}\right],\\
&= \sinh\left[\frac{d_p(t)}{2}\right] \approx \frac{d_p(t)}{2},
\end{split}
\end{equation}
which is as small as the adiabatic factors.

Note that, contrary to the time-independent case, the VDC $V_n^{(n)}(t) = \vg_0$ is not unique. In fact, any choice $V_n^{(n)}(t) = \vq$ in the region inside the disk with radius $d_\vq$ less then the minimum value of $d_p(t)/2$ during the time interval in which the adiabatic approximation is valid ($d_p^*/2$) yields approximately the same Bogoliubov coefficient given in \eqref{beta-distance-Ad}. As a concrete example, take as VDC an arbitrary point $V_n^{(n)}(t) = \vm_0^{(n)}$ (as defined in Eq.~\eqref{inicond}). For a solution $\vm^{(n)}$ emanating from $V_n^{(n)}(t)$, we obtain at first order in $d_p$
\begin{equation}
\label{beta-distance-Ad2}
\vert \beta_{V_n^{(n)},\vm^{(n)}}(t) \vert \lesssim \sinh(r) +\cosh(r)\frac{ d_p(t)}{2}.
\end{equation}
The expression above shows that if the VDC is close enough to the origin, i.e., $r \ll d_p$, then we have approximately the same distance from the VDC as in the case where $r=0$. In other words, any $r$ much smaller than $d_p$ give a VDC with approximately the same particle creation. Notwithstanding the result above, for simplicity, we will always choose the VDC as $V_n^{(n)} = \vg_0$.


Concluding, in the adiabatic case, we no longer have a simple stability curve (the curve which all solutions keep a constant distance from it), we now have a stability region, that is, the small circle around the origin with a radius smaller than $d_p$. Naturally, if $n<N_\mathrm{max}$, we can move to another adiabatic frame and reduce the size of the stability region. However, if the adiabatic series is asymptotic, then we have a finite $N_\mathrm{max}$ frame where the stability region attains its smallest size. Every curve $V(t)$ inside this region can be chosen, yielding a satisfactory vacuum state with negligible particle production during the time the adiabatic approximation is valid.


\subsection{The non-adiabatic case}
\label{subsec:V}

In the adiabatic case, the time scale in which the ``time''-like vector $\vu$ evolves is larger than the characteristic time of the frequency $\sigma$. In the previous sub-section, we took advantage of this fact to make canonical transformations to a frame where $\vu$ gets closer and closer to $\vg_0$. The crucial requirement for this is the fact that, after each transformation, the new vector $\vu^{(n)}$ is ``time''-like, and therefore it can be normalized. That is possible as long as the additional terms to the Hamiltonian vector (as $\delta \vu^{(1)}$ in Eq.~\eqref{u1}) have a norm smaller than one.

In the non-adiabatic cases, when we perform a canonical transformation to adiabatic frames, the transformed Hamiltonian is no longer ``time''-like. Moreover, in the literature, it is common to find physical situations described in a frame where the original Hamiltonian is already ``space''-like before any canonical transformation. For example, in the context of cosmological perturbations, the Mukhanov-Sasaki variable describes a harmonic oscillator with mass equal to one and frequency $\nu^2 = k^2 - V$ (where $k$ is the mode from the harmonic decomposition and $V$ the potential related to background quantities) in the conformal time gauge. When the potential $V$ gets larger than $k$, the frequency square is negative, the Hamiltonian is not positive definite anymore, and the vector $\vu$ is ``space''-like. Hence, as $M$ is necessarily ``time''-like (see Eq.~\eqref{M-g}), we cannot use the Hamiltonian vector $\vu$ or its corrections after performing canonical transformations to settle the initial conditions. Therefore, we have to find within this type of dynamics other ways to calculate an approximate propagator from which we can calculate the solutions and seek stability curves. Fortunately, this is possible in many physically relevant situations, as we will now see. 

After performing the canonical transformation \eqref{remove-h} to remove $h$,
the Hamiltonian vector $\vn$ is given by Eq.~\eqref{bar-n}:
\begin{equation}
\label{Ham-nonad}
	\vn = \left(\frac{1}{m}+m{\sigma}^{2}\right) \frac{\vg_0}{2} +  \left(\frac{1}{m}-m{\sigma}^{2} \right) \frac{\vg_2}{2}.
\end{equation}
In different physical scenarios, the non-adiabatic behavior takes place when $1/m$ or $m\sigma^2$, i.e., it diverges as $t\to t_0$. A common example is when $m$ is a positive power law in $t-t_0$ and $\sigma$ is real and regular at $t_0$, i.e.,
\begin{equation}\label{mpl}
\begin{split}
m &= {m_0}{(t-t_0)^\lambda}(1+\bigO{t-t_0}),\\ 
\sigma &= \sigma_0(1+\bigO{t-t_0}),
\end{split}
\end{equation}
where $m_0$, $\sigma_0$, $t_0$ are arbitrary real constants and $0<\lambda<2$ (the two special cases $\lambda=\pm2$ will be treated separately). This is an example where $N$ is ``time-like'', but the adiabatic approximation is not valid because $\dot{\xi}/(2\sigma) \propto \lambda/(t-t_0)$ becomes arbitrarily large as $t\to t_0$. Consequently, one must use another strategy. If it is possible to remove the $1/m$ term from the Hamiltonian \eqref{Ham-nonad} without adding more large terms, then the Hamiltonian vector becomes negligible in the limit $t\to t_0$. That would allow the usage of iterative approximations to calculate the associated dynamical propagator. Then, one can seek stability curves to which all the equations of motion solutions keep their distances almost constant.

To find such canonical transformations, it is convenient to write the Hamiltonian vector given in Eq.~\eqref{bar-n} as a linear combination of two null-vectors defined in Eq.~\eqref{defvl2} with time-dependent coefficients, i.e.,
\begin{equation}\label{hamu}
\vn =\frac{1}{m} \vl_2^+ -m\sigma^2\vl_2^- ,
\end{equation}
where the large $1/m$ term is isolated in front of $\vl_2^+$.
Using the properties of $\vl_2^{\pm}$, it is easy to see that the canonical transformation
\begin{equation}
\label{C+}
C_{+}(q) \equiv e^{-q \vl_2^+},
\end{equation} 
where $q$ is an arbitrary function of $t$,  leads to the Hamiltonian vector,\footnote{Such transformations (generated by $\vl_2^\pm$) when applied to the fields result in a redefinition of the field or momentum variable.}
\begin{align}
\label{transp}\vn^{(1)} &= C_{+}(q)\vn C^\dagger_+(q) + \dot{C}_+(q)C_+^\dagger(q),  \\
\nonumber  &= \vl_2^+\left(\frac{1}{m}+q^2m\sigma^2-\dot{q}\right) - \vl_2^-m\sigma^2+\vg_1 q m\sigma^2.
\end{align}
We can remove the term $1/m$ using the term $\dot{q}$. Otherwise, we would have to make $q^2 = -1/(m\sigma)^{2}$ which introduces a new large term through $\dot{q}$. In practice, we can solve the equation
\begin{equation}\label{eqq}
\dot{q} = \frac{1}{m}+q^2m\sigma^2,
\end{equation}
perturbatively. Considering the case where $m$ goes to zero as $(t-t_0)^\lambda$, i.e., the mass term is given exactly by $m=m_0(t-t_0)^\lambda$ for a constant $m_0$. The function
\begin{equation}
	\label{q0}
	q_0 \equiv \int\frac{1}{m}\dd t,
\end{equation}
is clearly a leading solution as long
\begin{equation}
\left\vert q_0^2m^2\sigma^2\right\vert \ll 1.
\end{equation}
Around $t_0$ the condition above translates to 
\begin{equation}
\left\vert\frac{(t-t_0)^2}{(\lambda-1)^2}\sigma_0^2\right\vert\ll1.
\end{equation}
There is always a $t$ close enough to $t_0$ such that this condition is satisfied. For a small correction $q=q_0+q_1$ and considering $q_1$ at the same order as $q_0^2m\sigma^2$, we obtain
\begin{equation}\label{q1}
\dot{q}_1 = q_0^2m\sigma^2, \qquad q_1 = \int q_0^2m\sigma^2\dd t.
\end{equation}
Again, this is a valid approximation as long
\begin{equation}
2q_1 \ll q_0.
\end{equation}
Then, around $t_0$ the condition is
\begin{equation}\label{condc}
\left\vert\frac{2(t-t_0)^2\sigma_0^2}{(\lambda-1)(\lambda-3)}\right\vert \ll 1.
\end{equation}
We can continue in this way and at each order we have a similar condition as~\eqref{condc}. In practice, this is the same result one would get solving this equation in powers of $m$ and considering $q_0$ a zero order term. Notwithstanding, that does not imply that $q_n$ are small, it is easy to show that around $t_0$, 
\begin{equation}\label{qseries}
q_n \propto (t-t_0)^{2n+1-\lambda}.
\end{equation}
Note that our previous imposition $\lambda<2$ makes $q_n$ for $n>0$ converges to zero at $t \to t_0$. This choice simplifies the analysis below, although it is not required. We could move on with $\lambda>2$, but the canonical transformations would be more complicated.

Substituting back this expression for $q$ up to first order in Eq.~\eqref{eqq}, we get 
\begin{equation}
\label{transpq}\vn^{(1)} = \vl_2^+\left(2q_0q_1 + q_1^2\right)m\sigma^2 - \vl_2^-m\sigma^2+\vg_1 q m\sigma^2,	
\end{equation}
where the first terms are at least second order in $m$ while the latter two are first order. Computing these quantities around $t_0$ results in the $\vl_2^+$ terms being the smallest (and the leading one proportional to $(t-t_0)^{4-\lambda}$), while the $\vl_2^-$ and $\vg_1$ are $(t-t_0)^{\lambda}$ and $(t-t_0)$ respectively. 

Now, considering only first order terms the Hamiltonian vector still has ``space-like'' components, $\vg_1$ and $\vg_2$. To get a Hamiltonian close to $\vg_0$, as we did in the adiabatic case, we need to eliminate such terms. For that, we make a first order canonical transformation generated by $\vg_1$ and $\vg_2$, i.e.,
\begin{equation}
\label{C11}
C_{12}(p_1,r_1) = e^{r_1\vg_2}e^{-  p_1 \vg_1},
\end{equation}
where $p_1$ and $r_1$ are arbitrary functions of time that are assumed to be first order in $m$. With this last transformation we get
\begin{equation}
\begin{split}
N^{(2)} &= \frac{m\sigma^2}{2}\vg_0+\left(q_0m\sigma^2-\dot{p}_1\right)\vg_1\\
&+\left(\dot{r}_1-\frac{m\sigma^2}{2}\right)\vg_2+\bigO{2}.
\end{split}
\end{equation}
Therefore, we can easily eliminate the terms proportional to $\vg_1$ and $\vg_2$ choosing,
\begin{equation}\label{p1r1}
p_1 = \int q_0 m\sigma^2\dd t,\qquad r_1 = \frac{1}{2}\int m\sigma^2\dd t,
\end{equation}
which are indeed first-order terms. Here we stress that we are considering $p_1$ and $r_1$ small compared to one since we are computing the canonical transformation in powers of these two functions, this is why choosing $\lambda<2$ simplifies the analysis. Studying the behavior around $t_0$, one can verify that there is always a value of $t$ close enough to $t_0$ such that these quantities are much smaller than one. However, note that these transformations add second order corrections to the $\vg_0$ term,\footnote{In addition to those already present in Eq.~\eqref{transpq}.} which we denote by $s_2$. Naturally, we can make a third canonical transformation of the kind $C_{12}(p_2,r_2)$ with second order functions $p_2$ and $r_2$ eliminating the same order terms in $\vg_1$ and $\vg_2$, again adding a correction term $s_3$ to $\vg_0$. Repeating this process up to order $n$ we get
\begin{equation}
N^{(n)} = s\vg_0+\bigO{n},
\end{equation}
where $s$ contains all correction terms up to order $n-1$, i.e.,
\begin{equation}
s = \frac{m\sigma^2}{2} + \sum_{i=2}^{n-1}s_i.
\end{equation}
In this frame, it is easy to compute the propagator up to $n-1$ order terms as
\begin{equation}
P^{(n)} = e^{\Delta\varphi\,\vg_0}, \qquad \Delta\varphi \equiv \int_{t_0}^{t} s\;\dd t.
\end{equation}
The solutions up to $\bigO{n}$ can be read from \eqref{M-static} by making the substitutions $\sigma \rightarrow s$, $\Delta\tau \rightarrow \Delta\varphi$, $U \rightarrow \vg_0$ and writing $M_0^{(n)}$ as in Eq.~\eqref{inicond}, we obtain
\begin{equation}\label{M-nonad}
\begin{split}
M^{(n)} &= \cosh r\,\vg_0 \\
&+ \sinh r \left(\cos(2\Delta\varphi)\bar{m}-\sin(2\Delta\varphi)\bar{m}\wedge\vg_0\right)\\
&+\bigO{n}.
\end{split}
\end{equation}
Similarly to the time-independent case, a general solution also evolves on circles in the ``spatial'' plane $\vg_1$, $\vg_2$. The crucial difference is that $|\Delta\varphi| \ll 1$, as it is a first order term. In this context, the solutions describe small arcs on the circles, while in the static and adiabatic cases they may spin around many times. Note that the Bogoliubov coefficient between $\vm_0^{(n)}$ and $\vm^{(n)}$ is
\begin{equation}
\left\vert\beta_{\vm^{(n)},\vm^{(n)}_0}\right\vert = \sin\vert\Delta\varphi\vert\sinh r+\bigO{n}.
\end{equation}
Consequently, as closer the initial conditions are from the origin, smaller is the arc segment length wandered by the solution. Therefore, the natural VDC choice in this frame is, again, $V^{(n)}(t)=\vg_0$. This is a stability curve in the same sense as before, the solutions do not change their distances from it as they evolve (up to the chosen order of approximation), yielding $\vert \beta_{M^{(n)},\vg_0}(t) \vert \approx \bigO{n}$. Of course, as in the adiabatic case, any other $V^{(2)}(t)$ with $r \leq \bigO{n}$ yields Bogoliubov coefficients of the same order. On the other hand, solutions with a very large $r$ will also have a very large $\beta$ coefficient, even with a first order $\Delta\varphi$. Concluding, at first order in $s$, the stability region is a small disc around the origin with a $\bigO{n}$ radius. Thus, the VDC for the non-adiabatic case is also defined by $V_n^{(n)} = \vg_0$, but now set in the non-adiabatic frames defined above.

\begin{figure*}	
	\centering \includegraphics[scale=1]{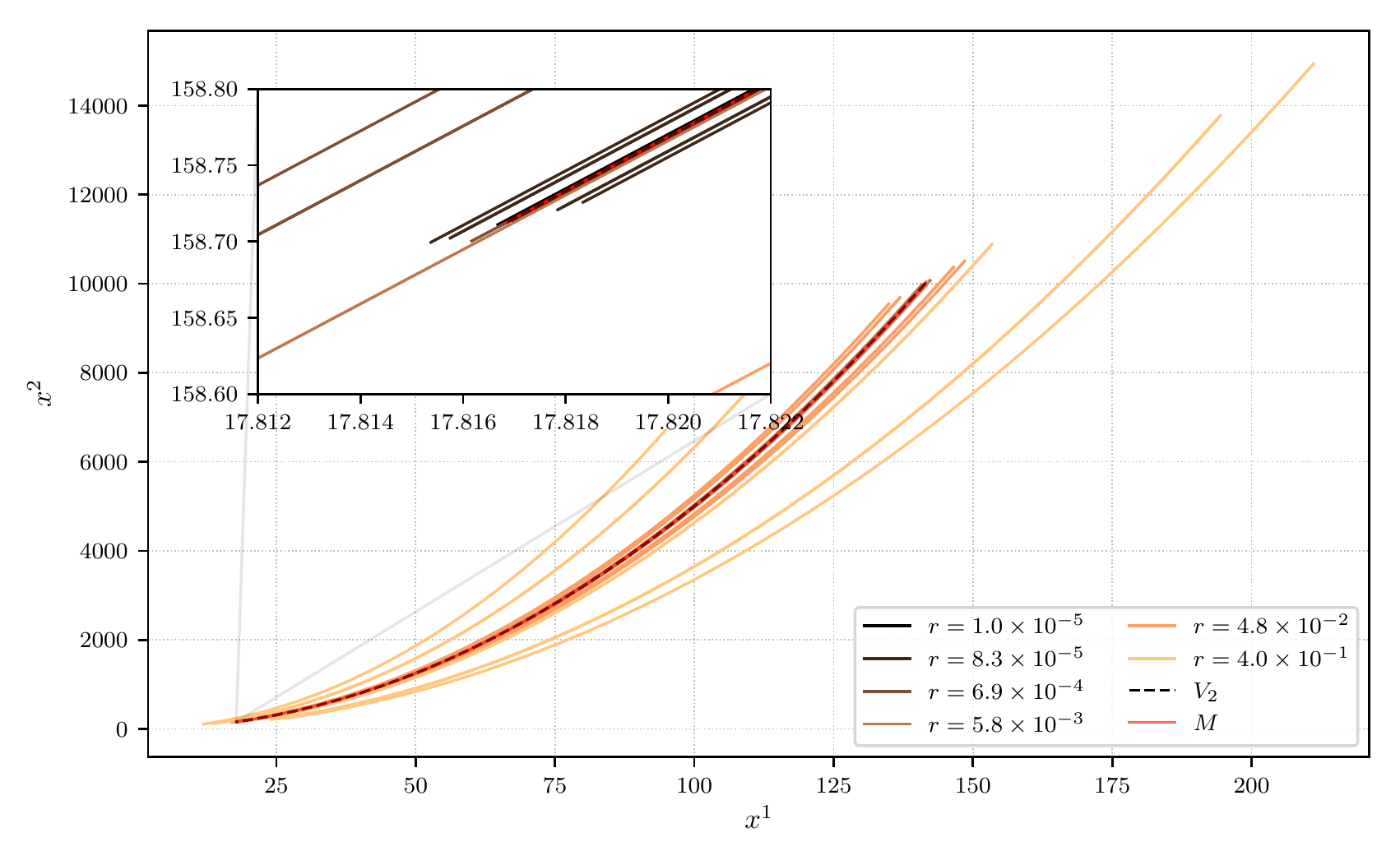} 
	\caption{\label{nonadiabf0}Numerical solutions for equation Eq.~\eqref{eqmovM} for the contracting model described in Sec.~\ref{bmcc}. All solutions were computed in the frame ${}^{(1)}$ and transformed to the original frame. The used parameters are $\Omega_w = 0.3$, $\Omega_\Lambda = 0.7$, $w=0.1$ and $k=10^3$. Since the solutions are ``time-like'' vectors with positive time components (as discussed below Eq.~\eqref{M-g}), the plotted ``spatial'' components $x^1$ and $x^2$ (see Eq.~\eqref{minkproj}) represent the solution uniquely. In this figure, we show solutions in the original frame. As discussed in Sec.~\ref{subsec:V}, this frame is dominated by the $1/m$ term, which can be removed using a canonical transformation generated by $q_0\vl_{2}^+$, leaving a first-order Hamiltonian vector. For this reason, the solutions in this frame for a given initial condition $M_0$ are given by Eq.~\eqref{approxq}, which draw approximately parabolas in the $x_1$--$x_2$ plane. We plot here $8\times6+1$ different solutions. The initial conditions (all set at frame ${}^{(1)}$) are given by points in circles around the origin $\vg_0$, with six equally log-spaced radius (hyperbolic distance from $\vg_0$) in the interval $(10^{-5}, 4\times10^{-1})$, and 8 equally spaced angles in $(0, 2\pi)$. We also plot the VDC given by $V_2(t)$ and the solution $\vm$ starting at $\vg_0$ and $T=0$, i.e., $M^{(1)}(0)=\vg_0$. In this frame, one cannot notice any discrepancy between them. Furthermore, the divergent term dominates the evolution, and all solutions diverge to infinity at $T=0$. Thus we only plot the interval $T \in (10^{-8}, 10^{-5})$. Solutions with the same distance to the VDC are plotted with the same color. We included a zoom at the end of evolution (near $T=0$) to present clearer the different solutions in this frame.}
\end{figure*}

\subsubsection{Special cases: $\lambda = 2$}
\label{secl2}

In the non-adiabatic regime there are also two special cases that can be solved in  another manner. Starting from the Hamiltonian~\eqref{hamu}, we can make the following canonical transformation
\begin{equation}
C_\pm(q_0)  = e^{\frac{\pi}{2}\vg_0}e^{\frac{1}{q_0}\vl_2^{-}}e^{-q_0\vl_2^{+}},
\end{equation}
where $q_0$ is the same function defined in Eq.~\eqref{q0}. The resulting Hamiltonian is
\begin{equation}\label{hamu2}
	\vn^{(1)} =\frac{1}{q_0^2m} \vl_2^+ -q_0^2m\sigma^2\vl_2^- .
\end{equation}
This transformation in the field perspective is just the re-scaling of $\phi$ by $1/q_0$. Considering the power-law behavior, when approaching the non-adiabatic regime given in Eq.~\eqref{mpl}, we have a new effective mass
\begin{equation}
	q_0^2m = \frac{1}{m_0} \frac{(t-t_0)^{2-\lambda}}{(\lambda-1)^2}.
\end{equation}
Thus, if  $\lambda = 2$, the new mass no-longer goes to zero when $t \to t_0$, i.e., $t_0$ is now a regular point in this frame. Moreover, as $m>0$ and assuming $\sigma>0$, then we can rewrite the Hamiltonian as
\begin{equation}
	\vn^{(1)} = \sigma e^{-\frac{\xi_q}{2}\vg_0\vg_2}\vg_0e^{\frac{\xi_q}{2}\vg_0\vg_2},\quad \xi_q \equiv \ln\left(q_0^2m\sigma\right).
\end{equation}
Making the inverse canonical transformation leads to another Hamiltonian as in Eq.~\eqref{adiabN1}, that is,
\begin{align}\label{scl2f2}
	\vn^{(2)} &= \sigma\left(\vg_0 - F_{q}\vg_1\right) ,  & F_{q} &\equiv -\frac{\dot{\xi}_q}{2\sigma}.
\end{align}
As $\xi_q$ is constant for $\lambda = 2$, we get just $\vn^{(2)} = \sigma\vg_0 $, as in the case examined in Sec.~\ref{constcase}., with an actual fixed point and the vacuum exactly determined by $V^{(2)}(t) = \vg_0$ in this frame. 

Going back to the original frame we obtain
\begin{equation}
V(t) =  \vl_2^+ \frac{q_0^2}{c_q} - \vl_2^-\left(\frac{1}{c_q} + \frac{c_q}{q_0^2}\right) - \vg_1 \frac{q_0}{c_q},
\end{equation}
where we defined the constant $c_q=q_0^2m\sigma$. Comparing the VDC above with the field variables as described by Eq.~\eqref{Mtofieldssol}, we obtain using $\BRi_a\doteq (\phi, \Pi_\phi)$, satisfying $\BRi_a^*\SM^{ab}\BRi_b = 1$ and a particular choice of time-dependent phase such that $\phi$ is real at the initial time
\begin{equation}
	\phi = \frac{q_0}{\sqrt{2c_q}}e^{-\ci\theta},\quad \Pi_\phi = \left(\frac{1}{\sqrt{2c_q}}-\ci \sqrt{\frac{c_q}{2}}\frac{1}{q_0}\right)e^{-\ci\theta},
\end{equation}
where 
\begin{equation}\label{eqtheta}
\theta=\theta_0 + \int\sigma\dd t.
\end{equation}
In this case not only we have an exact VDC, but it is also an exact solution.

\begin{figure*}	
	\centering \includegraphics[scale=1]{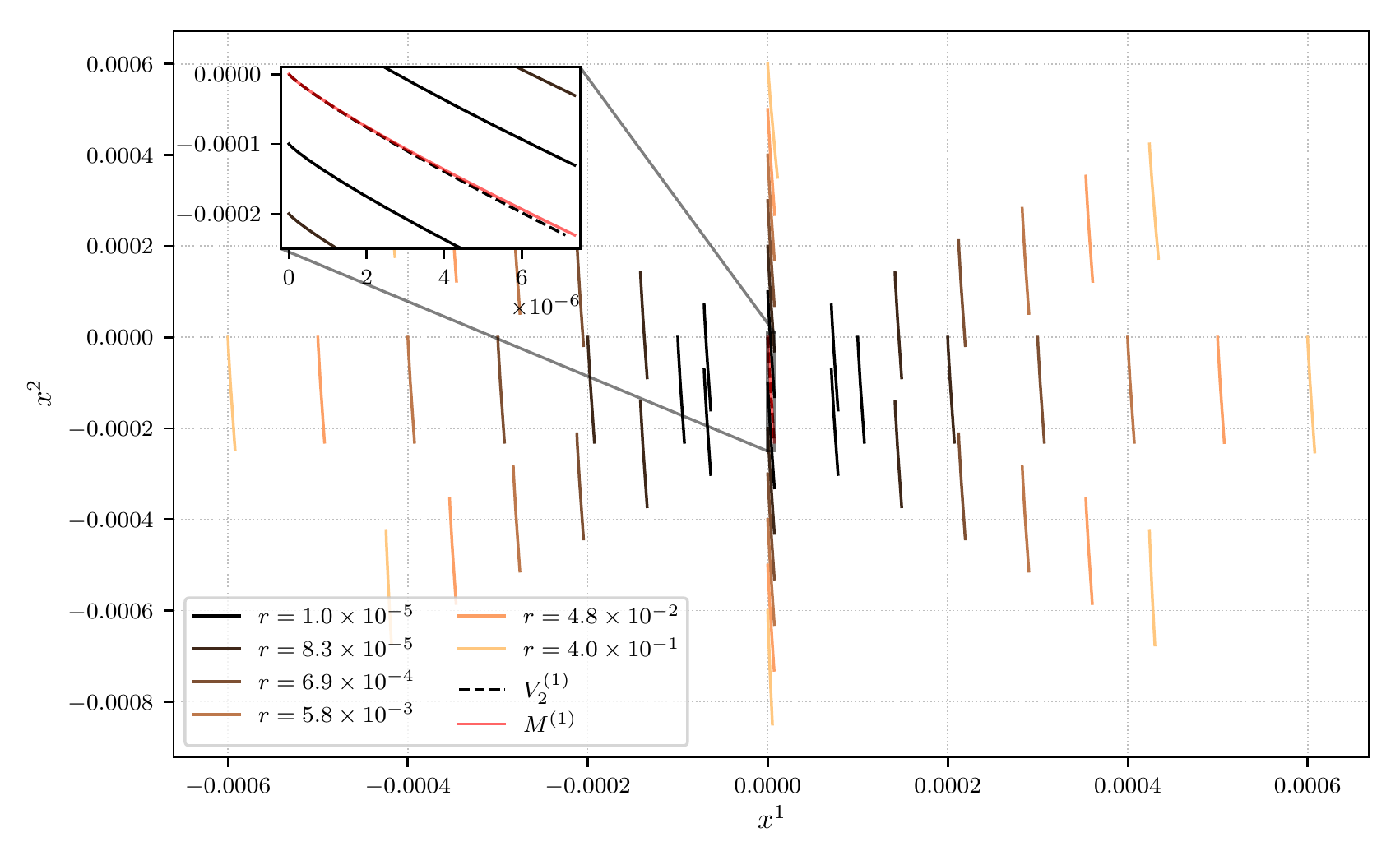} 
	\caption{\label{nonadiabf1} The same solutions discussed in Fig.~\ref{nonadiabf0} in the original frame now plotted in frame ${}^{(1)}$. The parabolic behavior seen in Fig.~\ref{nonadiabf0} has been removed by the canonical transformation. Consequently, we can see the first-order Hamiltonian effect on the solutions. The solutions at radii larger than the solutions' length are dragged towards the origin by applying a constant rigid motion using a re-scale function of the radius $f(r)$ as $(x^1(T), x^2(T)) \to (x^1(T), x^2(T)) - f(r)(x^1(0), x^2(0))$ to improve the readability of the plot. At the frame ${}^{(1)}$, the Hamiltonian vector~\eqref{transpq} has first-order terms in all components. For this reason, the solutions do not show any clear pattern as the evolution consists of a sequence of small boosts in different directions. Nevertheless, since all solutions keep an almost constant distance from the VDC, we notice that they follow similar trajectories as $V^{(1)}_2$. In this frame, the evolution of the VDC can be read directly from Eq.~\eqref{vdcp1r1} setting $q=0$, yielding $x^1\approx 2r_1$ and $x^2\approx 2p_1$. We also included a zoom of the VDC, $M^{(1)}$, and all nearby solutions. Here we can notice a small discrepancy between the VDC and $M^{(1)}$ at the end of the evolution.}
\end{figure*}

\subsubsection{Special cases: $\lambda=-2$}
\label{seclm2}

The second special case occurs when we have the same behavior for the mass~\eqref{mpl} but with $\lambda<0$. In these situations, the term $m$ must be removed from the Hamiltonian instead of $1/m$. To accomplish this, we need to perform a different canonical transformation generated by $\vl_2^-$ instead of $\vl_2^+$, 
Using $\vl_2^{\pm}$, the canonical transformation
\begin{equation}
	\label{C-}
	C_{-}(p) \equiv e^{p \vl_2^-},
\end{equation} 
where $p$ is an arbitrary function of $t$, leads to the Hamiltonian
\begin{align}
\vn^{(1)} &= \vl_2^+\frac{1}{m} + \vl_2^-\left(\dot{p}-m\sigma^2-\frac{p^2}{m}\right)-\vg_1\frac{p}{m}.
\end{align}
Similarly to the $1/m$ case we choose
\begin{equation}
	p = p_0,\qquad p_0 \equiv \int m\sigma^2\dd t.
\end{equation}
We could follow the same steps as in the beginning of this section to find the non-adiabatic vacuum. However, in this particular case, it is simpler to implement a second canonical transformation generated by $\vl_2^+$, 
\begin{equation}
	C_\mp(p_0)  = e^{\frac{\pi}{2}\vg_0}e^{-\frac{1}{p_0}\vl_2^{+}}e^{p_0\vl_2^{-}},
\end{equation}
to find the Hamiltonian
\begin{equation}\label{hamu3}
	\vn^{(1)} =\frac{p_0^2}{m} \vl_2^+ -\frac{m\sigma^2}{p_0^2}\vl_2^- .
\end{equation}
Note that, in this case the effective mass is given by
\begin{equation}
	\frac{m}{p_0^2} \approx  \frac{(\lambda+1)^2}{m_0\sigma_0^4}(t-t_0)^{-(2+\lambda)}.
\end{equation}
Thus, now the effective mass is constant when $\lambda = -2$. As in the previous case this Hamiltonian is a single boost in the $\vg_2$ direction, i.e.,
\begin{equation}
	\vn^{(1)} = \sigma e^{-\frac{\xi_p}{2}\vg_0\vg_2}\vg_0e^{\frac{\xi_p}{2}\vg_0\vg_2},\quad \xi_p \equiv \ln\left(\frac{m\sigma}{p_0^2}\right).
\end{equation}
Again, the second frame defined by the canonical transformation above leads to the VDC $V^{(2)}(t) = \vg_0$, which is a fixed point of the system when $m\propto t^{-2}$. In the original frame the VDC reads
\begin{equation}
	V(t) =  \vl_2^+\left(\frac{1}{p_0^2c_p}+c_p\right)  - \vl_2^- p_0^2c_p + \vg_1 p_0 c_p,
\end{equation}
where $c_p \equiv m\sigma/p_0^2$. Writing this VDC in the field variables using Eq.~\eqref{Mtofieldssol}), and a particular choice of time-dependent phase $\theta$ such that initially we have $\Pi_\phi$ purely imaginary, we get the normalized phase vector $\BRi_a\doteq (\phi, \Pi_\phi)$,
\begin{equation}\label{vdclm2}
	\phi = \left(\frac{1}{p_0\sqrt{2c_p}}+\ci\sqrt{\frac{c_p}{2}} \right)e^{-\ci\theta},\quad \Pi_\phi = -\ci\sqrt{\frac{c_p}{2}}p_0e^{-\ci\theta}.
\end{equation}
Here, $\theta$ is also given by Eq.~\eqref{eqtheta}.

\section{Applications}
\label{sec:Applications}

In this section we will apply the formalism developed in Sec.~\ref{sec:gcase} to some representative cases in Cosmology. For the adiabatic case, we will obtain the well-known adiabatic vacuum used to calculate the perturbation spectra in the matter bounce scenario. In the next sub-sections, we will apply the method constructed in Sec.~\ref{subsec:V} to two examples. We first consider the non-adiabatic case in which the contracting phase of a bouncing model is asymptotically dominated in the past by a cosmological constant. We obtain a new vacuum state for the cosmological perturbations which is not known in the literature. In the second case, in order to illustrate the power of the new method, we obtain the Bunch-Davies vacuum for a massless scalar field in de Sitter space in a completely different manner from those found in the literature. We use the VDC established in its appropriate frame as described in Sec.~\ref{sec:gcase}, and then we get back to the original frame in which the equations are usually formulated.

\begin{figure*}	
	\centering \includegraphics[scale=1]{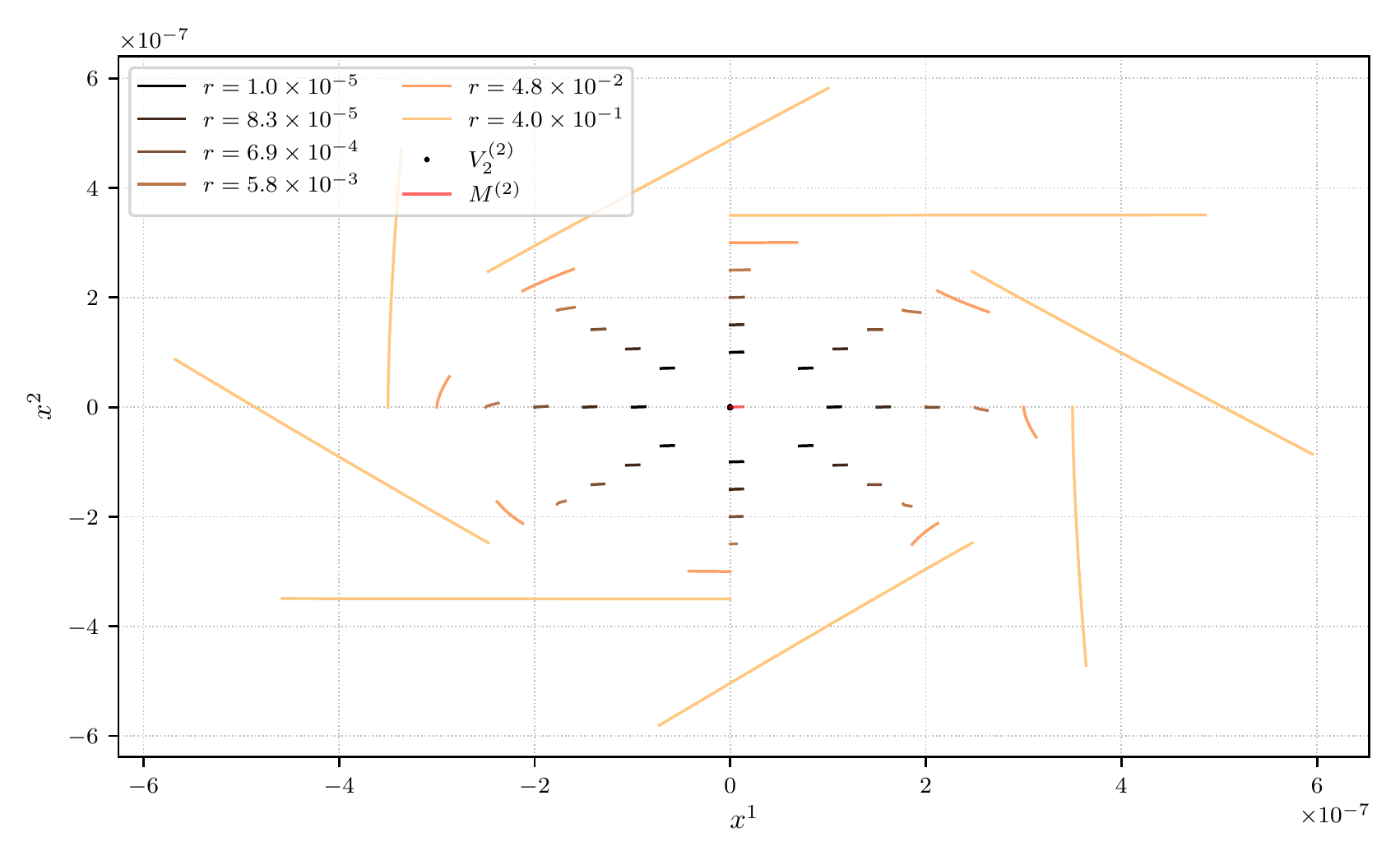} 
	\caption{\label{nonadiabf2} The same solutions presented in Figs.~\ref{nonadiabf0} and~\ref{nonadiabf1} are now plotted in frame ${}^{(2)}$. Since this frame removes the first-order terms from the Hamiltonian, the solutions (also moved by a rigid translation) present only second-order variations compared to Fig.~\ref{nonadiabf1}. We obtain the behavior predicted analytically in Eq.~\eqref{M-nonad}: solutions with large radii $r$ have their evolution dominated by the first order rotation describing arcs about $\vg_0$, which gets affected by second-order terms when the radius becomes sufficiently small. Since we are in a second-order frame, the VDC $V^{(2)}_2$ is given by the black dot at the origin, and the solution $M^{(2)}$ only deviates from it by second-order terms. This plot summarizes most of our results about the non-adiabatic VDC: it shows that all points close to the origin provide a reasonable VDC because they stay as close to their starting point as $M^{(2)}$ from $V_2^{(2)}$. That is the stability region discussed in Sec.~\ref{subsec:V}. As we move away from the origin, solutions start to rotate about $\vg_0$, and the second-order effects are less and less visible.}
\end{figure*}

\subsection{Non-adiabatic vacuum in bouncing models with a cosmological constant}
\label{bmcc}

One important application of the procedure described in Sec.~\ref{subsec:V} to provide a VDC is the problem of a stable quantum vacuum state definition in a bouncing model containing dark energy which is asymptotically dominated by it in the far past. In this limit of the contracting branch, the scalar and/or tensor perturbations field-mode frequencies are negligible with respect to the background expansion rate scale, that is, the mode-wavelengths are much larger than the Hubble radius at that time. Some attempts have been made~\cite{Maier2011} to define a sound quantum vacuum state in this situation. However, as discussed in Ref.~\cite{Maier2011}, their proposal had limitations, since it depends on a particular point during the contraction phase, where the potential is exactly zero and does not necessarily lead to a stable vacuum in the past. Note that if the standard $\Lambda$CDM cosmological model is correct and dark energy is a small cosmological constant, then any bouncing model containing a large scale contracting phase will have to face this problem, as the cosmological constant dominates at these scales. 

Consider a contracting Friedmann-Lema\^{\i}tre model dominated by a cosmological constant, and a perfect fluid with equation of state $p=w\rho$, for a constant $w$, where $p$ is the fluid pressure and $\rho$ its energy density. In bouncing models one usually takes $|w| \ll 1$, the so-called matter bounce, as it naturally leads to an almost scale invariant spectrum of scalar perturbations when the matter fluid dominates the contracting phase. Let us define the redshift time function $x=a_0/a$, where $a$ is the scale factor and $a_0$ is its value today. In this situation, the Hamiltonian of scalar linear perturbations has mass and frequency terms given by~\cite{Vitenti2013}
\begin{equation}
\label{fm}
\sigma = \lapse x c_s k, \quad m = \frac{4\pi G(\rho + p)}{\lapse c_s^2 x^3 H^2},
\end{equation}
where $\lapse$ is the lapse function, $G$ is the Newton constant, $c_s$ is the sound velocity of the scalar perturbations, $H$ is the Hubble function and $k$ is the wave-number.  Dynamics in the far past of the contracting phase is ruled by classical General Relativity, yielding the Friedmann equation
\begin{equation}
\label{Friedmann}
\frac{H^2}{H_0^2} = \Omega_w x^{3(1+w)} + \Omega_\Lambda,
\end{equation}
where $H_0$ is the Hubble constant today, $\Omega_w \equiv \rho_0/\rho_{c0}$, $\Omega_\Lambda \equiv \Lambda/(8\pi G\rho_c)$, $\rho_0$ is the fluid's energy density today, and $\rho_{c0} \equiv 3H^2_0/(8\pi G)$ is the critical density today.

Using the dimensionless time parameter,\footnote{From now on, in this subsection and in the following one, all variables are made dimensionless as they express physical quantities in Hubble radius unities, that is, $t_S \equiv t H_0$, $k_S = (1/H_0)k$, \emph{etc}.} $0<T<\infty$ defined by $T \equiv x$ (this time variable is convenient since it tends to the conformal time in the asymptotic past), yields
\begin{equation}
\lapse=\frac{1}{T\sqrt{\Omega_\Lambda}}\frac{1}{\sqrt{1+\alpha T^{2+\lambda}}},
\end{equation}
where $\alpha \equiv \Omega_w /\Omega_\Lambda$ and $\lambda\equiv 1+3w$. Substituting this time variable back in~\eqref{fm}, results in
\begin{equation}
\label{fm1}
\sigma = \frac{\sigma_0}{\sqrt{1+\alpha T^{2+\lambda}}},\qquad m = \frac{m_0 T^{\lambda}}{\sqrt{1+\alpha T^{2+\lambda}}},
\end{equation}
where 
$$\sigma_0 \equiv \frac{c_s k}{\sqrt{\Omega_\Lambda}},\qquad  m_0 \equiv\frac{(2+\lambda)\alpha\sqrt{\Omega_\Lambda}}{2c_s^2} ,$$

We are interested in the asymptotic  past of the contracting phase, when $T \to 0$. In this case one gets
\begin{equation}
\label{fm2}
\sigma \approx \sigma_0,\qquad m \approx m_0 T^{\lambda}.
\end{equation}
Using Eq.~\eqref{fm2} in Eqs.~(\ref{q0}, \ref{q1}, \ref{p1r1}) we get
\begin{align}\label{plq0q1}
q_0 &=  \frac{T^{1-\lambda}}{m_0(1-\lambda)}, & q_1 &= \frac{T^{3-\lambda}\sigma_0^2}{m_0(3-\lambda)(1-\lambda)^2},\\ 
\label{plp1r1}
p_1&=\frac{T^{2}\sigma_0^2}{2(1-\lambda)}, & r_1 &= \frac{T^{1+\lambda}m_0\sigma_0^2}{2(1+\lambda)}.
\end{align}
Then, the VDC discussed in Sec.~\ref{subsec:V} is just the choice $V^{(2)}_2(t)=\vg_0$ in the transformed frame ${}^{(2)}$.

Performing the inverse transformations we obtain the VDC in the original frame. In the case of a matter bounce scenario where $0<w\ll 1$ one gets that $\lambda$ is slightly larger than one. For this reason, we see that $q_0$ is actually divergent when $T\to0$, while all other transformation parameters are small in the limit. For this reason, we first write the full transformation as
\begin{equation}\label{nonVDCfull}
	\begin{split}
	&\vv_2(t) = C_{+}^\dagger(q) C_{12}^\dagger(p_1,r_1)\vg_0 C_{12}(p_1,r_1) C_{+}(q),\\
	&=\cosh(2r_1)\left[\cosh(2p_1)+\frac{q^2}{2}e^{-2p_1}-q\tanh(2r_1)\right]\vg_0 \\
	&+\cosh(2r_1)\left[-qe^{-2p_1}+\tanh(2r_1)\right]\vg_1\\
	&+\cosh(2r_1)\left[\sinh(2p_1)+\frac{q^2}{2}e^{-2p_1}-q\tanh(2r_1)\right]\vg_2.
	\end{split}
\end{equation}
Expanding $\vv_2(t)$ at first order in $p_1$ and $r_1$, we get
\begin{equation}\label{vdcp1r1}
	\begin{split}
		\vv_2(t) &=\left(1+\frac{q^2}{2}-2qr_1-q^2p_1\right)\vg_0 \\
		&+\left(2r_1-q+2qp_1\right)\vg_1\\
		&+\left[\frac{q^2}{2}-2qr_1+\left(1-\frac{q^2}{2}\right)2p_1\right]\vg_2.
	\end{split}
\end{equation}

We can relate the full VDC~\eqref{nonVDCfull} with the field variables using Eq.~\eqref{Mtofieldssol}, yielding
\begin{align}
\phi &= q e^{-p_1}\frac{1-\ci\sinh(2r_1)}{\sqrt{2\cosh(2r_1)}}+\ci e^{p_1}\sqrt{\frac{\cosh(2r_1)}{2}},\\
\Pi_\phi &= e^{-p_1}\frac{1-\ci\sinh(2r_1)}{\sqrt{2\cosh(2r_1)}}.
\end{align}
At first order, the same VDC reads,
\begin{align}
\phi &=\frac{q(1-p_1)+\ci(1+p_1-2q r_1)}{\sqrt{2}}, \\
\Pi_\phi &= \frac{1-p_1+2\ci r_1}{\sqrt{2}}.
\end{align}
For our present problem, we can express this VDC explicitly in terms of the cosmological parameters, which reads
\begin{align}
\nonumber
&\phi \approx \frac{\ci}{\sqrt{2}}\left(1-\frac{c_s^2 k^2 T^2}{2\Omega_{\Lambda}(2+3w)}\right) \\
\label{super-Hubbleftest2}&+ \frac{1}{\sqrt{2}}\left[-\frac{2c_s^2\sqrt{\Omega_{\Lambda}}T^{-3w}}{9w(1+w) \Omega_{w}} \left(1-\frac{c_s^2 k^2 T^2}{2\Omega_{\Lambda}(2-3w)}\right)\right],\\ \nonumber
&\Pi_{\phi} \approx \frac{\ci}{\sqrt{2}}\left(-\frac{3(1+w) \Omega_{w} k^2 T^{2+3w}}{2(\Omega_{\Lambda})^{3/2}(2+3w)}\right)\\
\label{super-Hubbleptest2}&+\frac{1}{\sqrt{2}}\left(1+\frac{c_s^2 k^2 T^2}{6w\Omega_{\Lambda}}\right).
\end{align}
The limit $T\to 0$ corresponds to the super-Hubble regime, hence the field variables $\phi$ can be obtained using the super-Hubble approximation at the same order, that is,
\begin{equation}
\begin{split}
	\phi &\approx A_1(k) \left(1-\int\frac{\dd T}{m}\int m'\sigma^{\prime 2} \dd T'\right)\\
	&+ A_2(k) \left(\int\frac{\dd T}{m}-\int\frac{\dd T}{m}\int m'\sigma^{\prime 2} \dd T' \int\frac{\dd T''}{m''}\right),
\end{split}
\end{equation}
where the primed functions indicate the argument, e.g., $m'\equiv m(T')$. One can verify that the time-dependence calculated in this way fits with Eq.~\eqref{super-Hubbleftest2}, leading to the particular choice $A_1(k) = \ci A_2(k) = \ci/\sqrt{2}$ (apart from a $k$-independent constant phase). Therefore, these two constants, which yield the spectrum of $\phi$, are determined using our VDC. 

Equations~\eqref{super-Hubbleftest2} and \eqref{super-Hubbleptest2} are the main result of this subsection. It yields a new stable vacuum state, which can be defined in the asymptotic contracting phase of a bouncing model with a cosmological constant and a matter fluid, yielding the initial conditions of quantum cosmological perturbations on it. It is worth emphasizing that this result is valid for any physical case where only $q_0$ diverges in the limit $T\to0$. 

The behaviors described in this section can be seen by examining Figs.~\ref{nonadiabf0}, ~\ref{nonadiabf1} and~\ref{nonadiabf2}. In these figures we use the Minkowski representation where the coordinates $x^1$ and $x^2$ are ``spatial'' projections and $x^0$ is the ``time'' component, i.e., 
\begin{equation}\label{minkproj}
x^B \equiv \vm\cdot\vg_A\eta^{AB}.
\end{equation}
The original frame shown in Fig.~\ref{nonadiabf0} is dominated by the $q$ term, this is easily seen inspecting an arbitrary solution
\begin{align}
\nonumber \vm &=C_{+}^\dagger(q)P^{(1)}M_0P^{(1)\dagger}C_{+}(q),\\
\label{approxq}&= C_{+}^\dagger(q)M_0C_{+}(q) + \bigO{1},
\end{align}
In terms of the components we have at leading order, ignoring first order corrections
\begin{align}
x^1 &= x^1_0 +q(x^1_0+x^0_0), \\
x^2 &= x^2_0- q x^1_0 -\frac{q^2}{2}(x^2_0+x^0_0) \\
x^0 &= x^0_0+ q x^1_0 +\frac{q^2}{2}(x^2_0+x^0_0).
\end{align}
For a given initial condition components $x_0^B = \vm_0\cdot\vg_A\eta^{AB}$. Thus, we have approximately parabolas in the $x^1$--$x^2$ plane.

The first order frame is depicted in Fig.~\ref{nonadiabf1}, where the divergent term is removed. Finally, in the second order frame (Fig.~\ref{nonadiabf2}), we see the rotating behavior dominating far from the VDC. As we get closer to the origin, the solutions follow the behavior of the VDC.

\begin{figure*}	
	\centering \includegraphics[scale=1]{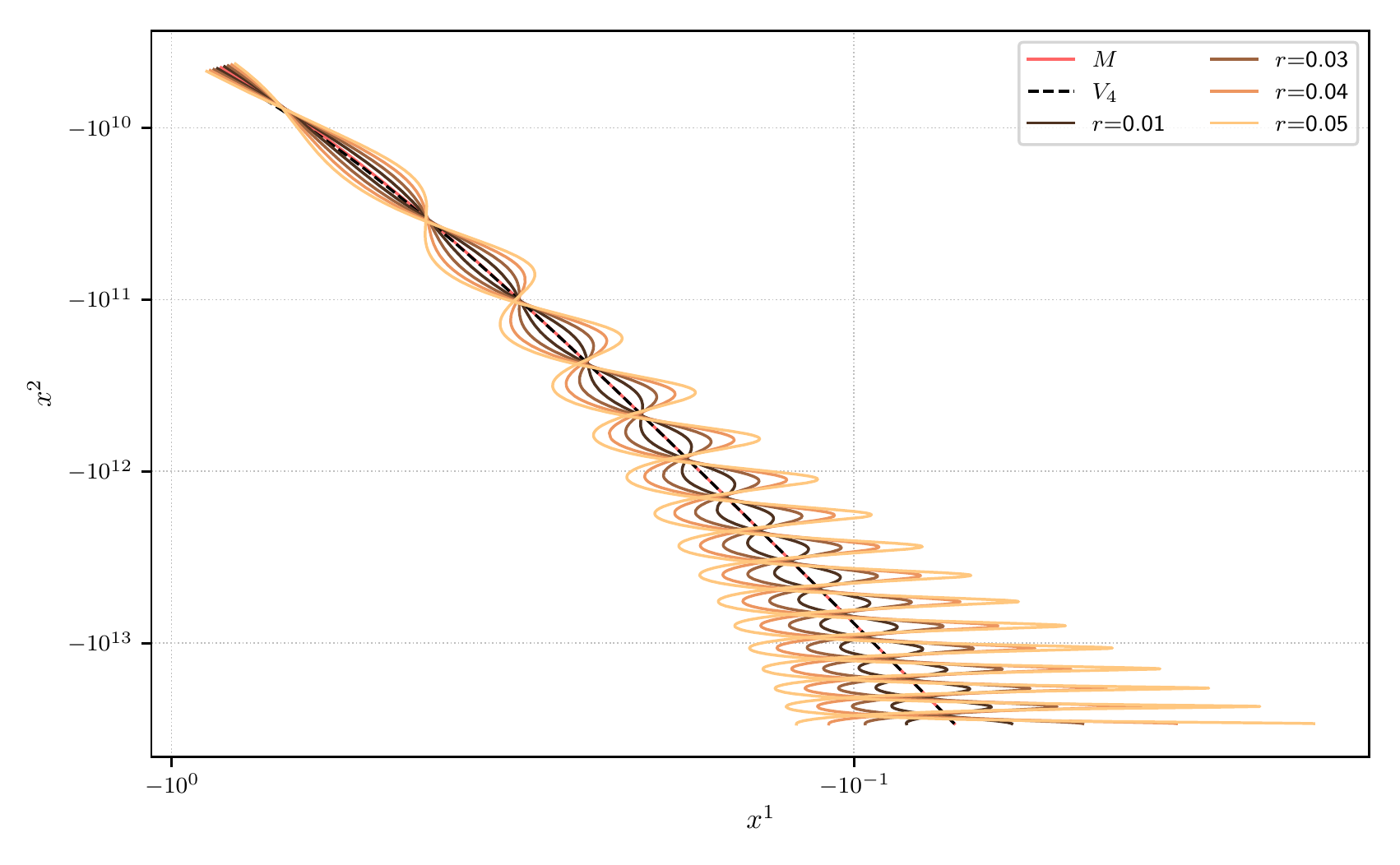} 
	\caption{\label{adiabf0} Numerical solutions for Eq.~\eqref{eqmovM} for the contracting model described in Sec.~\ref{sec:adiab}. All solutions were computed at the first adiabatic frame ${}^{(1)}$. The used parameters are $\Omega_w = 1$, $w=10^{-6}$ and $k=10^2$. As in Fig.~\ref{nonadiabf0}, we plotted the components $x_1$ and $x_2$ representing a solution uniquely. In this figure, we show the solutions in the original frame. As discussed in Sec.~\ref{sec:adiab}, this frame is boosted by $\xi$. Consequently, the VDC and solutions in this frame move quickly in the $\vg_2$ direction. We plot here $4$ different solutions in addition to the VDC, and the solution $M$ satisfying $M(T_0) = V_4(T_0)$ where $T_0=10^{-10}$. Initial conditions are given by circles around the VDC with distances from $\vv_4$ for $r=0.01, 0.03, 0.04$ and $0.05$, and the two angles $0$ and $\pi$. Solutions with the same distance to $\vg_0$ are plotted with the same color.}
\end{figure*}

\begin{figure*}	
	\centering \includegraphics[scale=1]{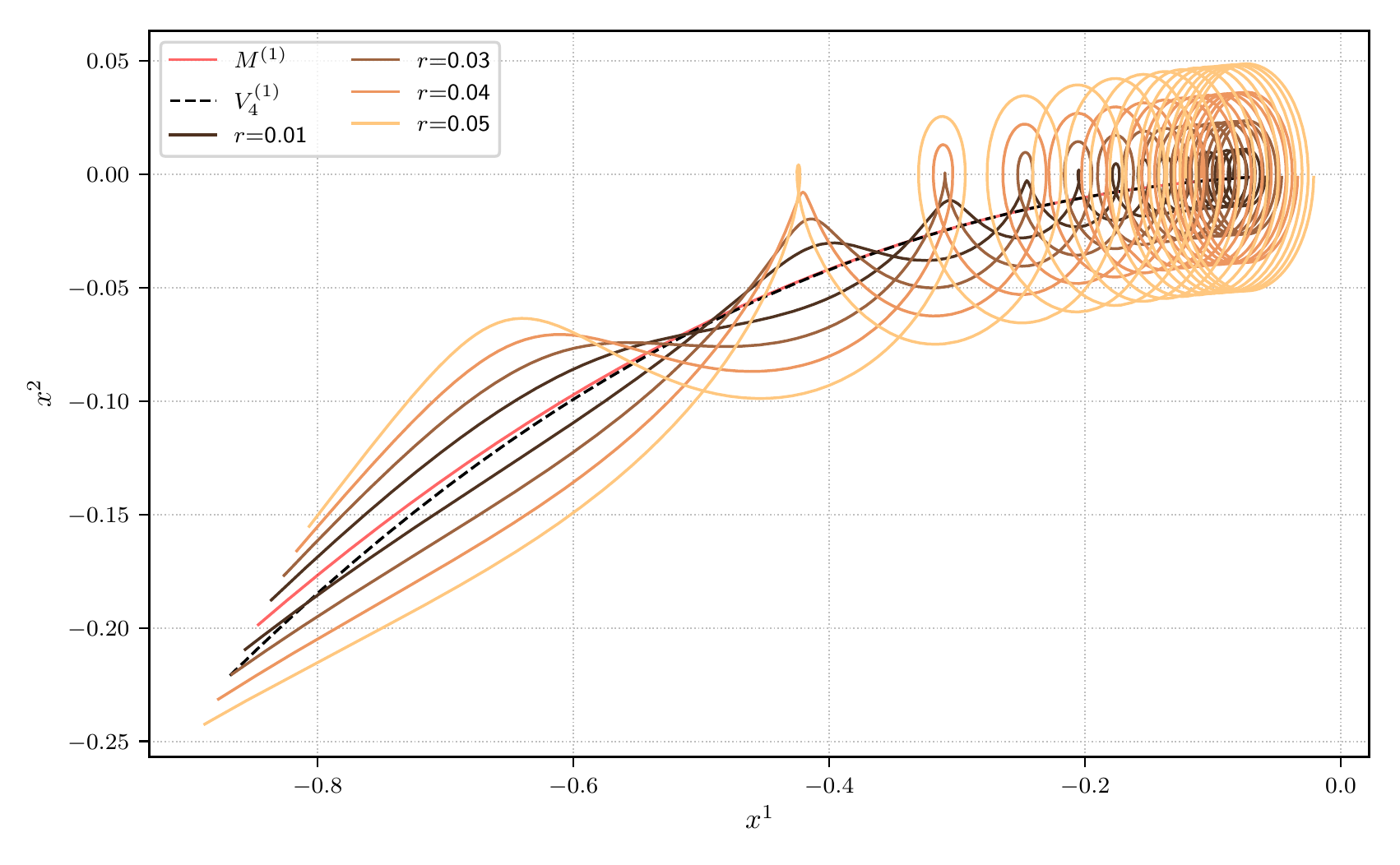} 
	\caption{\label{adiabf1} The solutions presented in Fig.~\ref{adiabf0} are plotted here in the first adiabatic frame. This frame removes the boost generated by $\xi$. Consequently, the motion in the $\vg_2$ direction is reduced. We predicted this behavior analytically. As we increase the index $(n)$ of the adiabatic frame, the VDC stays closer to the origin, and solutions keep rotating around it.}
\end{figure*}

\begin{figure*}	
	\centering \includegraphics[scale=1]{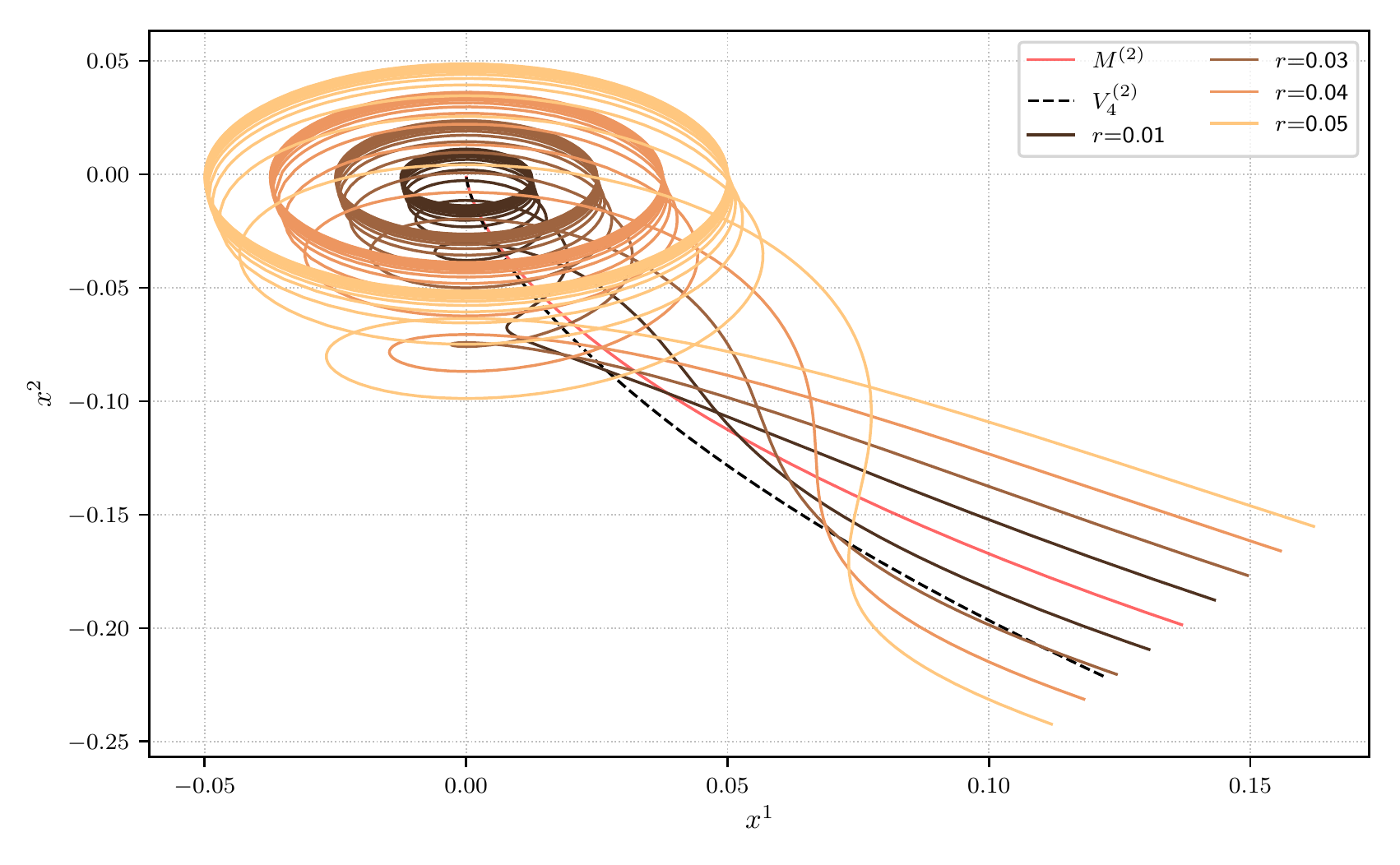} 
	\caption{\label{adiabf2} The same solutions presented in Figs.~\ref{adiabf0} and Fig.~\ref{adiabf1} are plotted in the second adiabatic frame. The boosts generated by $\xi$ and $\xi_1$ are removed. Therefore both movements in the $\vg_2$ and $\vg_1$ directions are reduced. Note that the VDC and $M^{(2)}$ begin to separate only when the adiabatic condition is no longer fulfilled. We can visualize this as the region where the solutions cease to revolve around the stability curve.}
\end{figure*}

\subsubsection{Special case: radiation}

For a radiation fluid we have $w=1/3$, hence $\lambda = 2$. This is the case described in Sec.~\ref{secl2}. At leading order in $T$ (see Eq.~\eqref{fm2}) we have 
\begin{equation}
c_q = \frac{\cs^3k}{2\alpha\Omega_\Lambda}, \qquad q_0 = -\frac{\cs^2}{2T\alpha\sqrt{\Omega_\Lambda}}.
\end{equation}
Using these quantities we obtain the VDC in terms of the fields as
\begin{equation}
	\phi = -\sqrt{\frac{\cs}{k\alpha}}\frac{1}{2T}e^{-\ci\theta},\quad \Pi_\phi = \sqrt{\frac{\alpha}{\cs k}}\left(\frac{\sqrt{\Omega_\Lambda}}{\cs}+\ci kT\right)e^{-\ci\theta}.
\end{equation}
Note that in this case the field diverges as $T\to0$ in the original frame. However, even in this case we have a well-defined vacuum.

\subsection{Adiabatic vacuum in the matter bounce scenario}
\label{sec:adiab}
In this section we make a brief review of the adiabatic vacuum in the matter bounce model in order to illustrate our method in a well-known scenario. Here we use the same model as in Sec.~\ref{bmcc}, but setting $\Omega_\Lambda=0$. We them get the following mass and frequency results:
\begin{equation}
	\sigma = \frac{\sigma_0}{ T^{1+\lambda/2}},\qquad m = \frac{m_0}{T^{1-\lambda/2}},
\end{equation}
where here we have
$$\sigma_0 \equiv \frac{c_s k}{\sqrt{\Omega_w}},\qquad  m_0 \equiv\frac{(2+\lambda)\sqrt{\Omega_w}}{2c_s^2}.$$
Substituting these quantities in Eq.~\eqref{adiabN1} results in
\begin{equation}
	F_1 = \frac{\sqrt{\Omega_w T}}{\cs k},
\end{equation}
which tends to zero as $T\to0$. One can compute higher order terms to check that all converge to zero as $T\to0$.

To illustrate our method, in Fig.~\ref{adiabf0} we plot the solutions calculated numerically using a fourth order adiabatic vacuum as the VDC in the original frame. We computed both $F_1$ and $F_2$ analytically and $F_3$, and $F_4$ numerically. However, the numerical results provide no further insights on the problem, and therefore we are not including them here.\footnote{The numerical code and python notebooks where these solutions were computed can be found here: \href{https://github.com/NumCosmo/NumCosmo/blob/master/notebooks/VacuumStudy.ipynb}{VacuumStudy} and \href{https://github.com/NumCosmo/NumCosmo/blob/master/notebooks/VacuumStudyAdiabatic.ipynb}{VacuumStudyAdiabatic}.} We also included solutions starting around the VDC with a distance $r$ and angle $\theta$ with the $x^1$ axis. Although the frame is dominated by the boost generated by $\xi$, we can see that the solutions oscillate around the VDC. In Fig.~\ref{adiabf1} we show the same solutions in the first adiabatic frame. In this frame, the $\xi$ boost is removed, and the solutions are closer to circles around the origin. Finally, in Fig.~\ref{adiabf2} we plot again the same solutions but now in the second adiabatic frame. Here we notice a further slow-down of the VDC, as it stays even closer to the origin than in the first frame. Solutions starting around the VDC have the predicted behavior, they revolve and move keeping its center on the VDC. Finally, in all plots we also included a solution starting exactly at the VDC at the initial time $T_0 = 10^{-10}$, note that it stays initially close to the VDC, and starts to deviate from it by the end of the adiabatic regime.

\subsection{Bunch-Davies vacuum in the non-adiabatic formalism}

The Hamiltonian vector \eqref{hamu}, 
\begin{equation}\label{hamu4}
\vn =\frac{1}{m} \vl_2^+ -m\sigma^2\vl_2^- ,
\end{equation}
which sets the dynamics in conformal time $\eta$ for the modes $\phi_k(\eta)$ of a free scalar field with mass $M$ in de Sitter space have the functions $m(\eta)$ and $\sigma(\eta)$ given by
\begin{equation}
\label{msigma-dS}
m(\eta)=a^2(\eta)=\frac{1}{H_{\Lambda}^2\eta^2},\quad \sigma^2(\eta) = k^2 + M^2 m(\eta),
\end{equation}
where $H_{\Lambda}^2\equiv \Lambda/3$, $a(\eta)$ is the scale factor of de Sitter space in flat coordinate system, and $\Lambda$ is the cosmological constant.

From the Hamiltonian vector defined by Eqs.~\eqref{hamu2} and \eqref{msigma-dS}, we can obtain the following second order equation for the mode $v_k(\eta)\equiv \sqrt{m(\eta)}\phi_k(\eta)$, namely
\begin{equation}
    \frac{{\rm d}^2v_k}{{\rm d}\eta^2}+\left[k^2-\left(2-\frac{M^2}{H_{\Lambda}^2}\right)\frac{1}{\eta^2}\right]v_k=0\,,
    \label{scalar-dS}
\end{equation}
whose general solution reads
\begin{equation}
\label{sol-sdS}
v_k (\eta) = \sqrt{k|\eta|}\left[A_1(k) J_{\gamma}(k |\eta|) + A_2(k) Y_{\gamma}(k |\eta|)\right].
\end{equation}
where $$\gamma \equiv \sqrt{\frac{9}{4} - \frac{M^2}{H_{\Lambda}^2}},$$ and
$J_{\gamma}$ and $Y_{\gamma}$ are the Bessel functions of first and second kind, respectively.

One common way to fix $A_1(k)$ and $A_2(k)$ and get the Bunch-Davies vacuum is to impose the adiabatic vacuum prescription in the period where the adiabatic approximation is valid. Take the first adiabatic function (see App.~\ref{app:adiab}) of this model, 
\begin{equation}
\label{F1-BD}
F_1=\frac{1}{2\sigma}\frac{{\rm d}\xi}{{\rm d}\eta}=-\frac{H_{\Lambda}}{2k}\frac{2k^2 H_{\Lambda}^2 \eta^2+3M^2}{(k^2 H_{\Lambda}^2 \eta^2+M^2)^{3/2}},
\end{equation}
where $\xi=\ln(m\sigma)$. When $|k\eta|\to\infty$, which is also the ultraviolet limit, $|F_1|\to 0$, as well as all others $|F_n|$, and the adiabatic approximation is applicable in this regime. Expanding the Bessel functions in the limit $|k\eta|\to\infty$, solution \eqref{sol-sdS} reduces to the adiabatic vacuum
iff $A_1(k) = \ci A_2(k) = (\pi/(4k))^{1/2}$, yielding the Bunch-Davies vacuum mode,
\begin{equation}
\label{Bunch-Davies}
v_k (\eta) = \frac{\sqrt{\pi|\eta|}}{2}\left[J_{\gamma}(k |\eta|) - \ci Y_{\gamma}(k |\eta|)\right].
\end{equation}
For a fixed time $\eta$ one need to take the limit $k \to \infty$ to define the vacuum state. That is, the behavior of $A_1(k)$ and $A_2(k)$ is only defined in this limit. Therefore, one must study the behavior of the solutions at different time frames to get  this behavior for all values of $k$.

In this subsection we will obtain the Bunch-Davies vacuum exploring the features of the Hamiltonian vector \eqref{hamu2} outside the adiabatic region, where $|\eta|\to 0$, which is the super-Hubble regime, using the method constructed in Sec.~\ref{subsec:V}. We will focus on the massless case.

In the massless case $M=0$ we have $\gamma = 3/2$, $\sigma = k$, $m$ is the same, and the $F_1$ function reads
\begin{equation}
\label{F1-BD-massless}
F_1=\frac{1}{2k}\frac{{\rm d}\xi}{{\rm d}\eta}=-\frac{1}{k \eta}.
\end{equation}
The expression above shows that in the limit $\eta\to0$ the adiabatic prescription cannot be used. However, this is exactly the special case discussed in Sec.~\ref{seclm2}, compare the mass term given in Eq.~\eqref{msigma-dS} with Eq.~\eqref{mpl}. Therefore, substituting $m$ and $\sigma$ in Eq.~\eqref{vdclm2} we obtain the following VDC applied to the fields,
\begin{equation}\label{bdvdc}
	\phi = -\frac{1}{a}\frac{1}{\sqrt{2k}}\left(1-\frac{\ci}{ k\eta}\right)e^{-\ci\theta},\quad \Pi_\phi = \ci a\sqrt{\frac{k}{2}}e^{-\ci\theta}.
\end{equation}
This is the well-known Bunch-Davies vacuum for a massless scalar field with minimal coupling.

Some remarks are important here. Note that we have obtained the Bunch-Davies vacuum Eq.~\eqref{bdvdc} in the massless case from the Hamiltonian Eq.~\eqref{hamu4} directly. All canonical transformations were used without any approximation in all steps, hence the final Hamiltonian vector~\eqref{hamu3} is constant. We are not aware of any attainment of the Bunch-Davies vacuum where the general solution \eqref{sol-sdS} is not used, and through an inspection of its large scale modes behavior alone.

The route we have used to arrive at Eq.~\eqref{hamu3} is not useful in the other cases, since it does not lead to the intermediate Hamiltonian with constant coefficients for other time dependencies. Thus, the last canonical transformation that takes the Hamiltonian to $\vg_0$ will insert back in the Hamiltonian vector an extra term with $\vg_1$ multiplied by the time-derivative of $\xi_p$. As the reader may verify, this is also the case of the massive scalar field in de Sitter space, where time-dependent terms appear in the intermediate steps.

\section{Conclusion}
\label{sec:conclusion}

In this paper, we built a framework to find stable vacuum solutions in terms of stability points in the hyperbolic space $\mathbb{H}^2$ defined as the points to which all solutions of the mode equations mapped to this space keep approximately the same distance from them. When the mode frequencies dominate the field mode evolution, we recover the usual adiabatic vacua in simple terms. The method is then applied to the reverse physical situation, namely, when the mode frequencies become irrelevant for the field mode evolution or, in the case of curved spaces when the mode wavelengths are larger than the curvature scale of the background space-time. We show that these stability points can also be found, yielding reasonable stable vacuum states. 

As an illustrative example of the second method, we obtained the Bunch-Davies vacuum of de Sitter space in a completely different manner. We carried out the analysis in the super-Hubble regime, where the adiabatic approximation is not applicable. In this regime, we showed that the time-dependent Hamiltonian vector in the hyperbolic space $\mathbb{H}^2$ describing the dynamics of a massless scalar field in the de Sitter space can be exactly transformed to a trivial time-independent Hamiltonian using boosts and rotations in this space. This simple Hamiltonian vector yields its obvious vacuum state that, when transformed back to the original frame, is exactly the Bunch-Davies vacuum of the massless scalar field in de Sitter space. In all this analysis, we never used the general solution of the field mode equations. To the best of our knowledge, this is the first time one obtained the Bunch-Davies vacuum without recurring to ultraviolet limit or the general solution~\eqref{sol-sdS}.

We then applied the new method to the cosmological problem of finding the vacuum state of cosmological perturbations in bouncing models with a cosmological constant and a fluid satisfying $p/\rho = w= \text{const}$ in the asymptotic past of such models. In this regime, the mode frequencies are irrelevant for the mode dynamics, and all relevant cosmological scales are much bigger than the Hubble radius. Nevertheless, we found a new vacuum state, given in Eq.~\eqref{vdcp1r1}, which is stable up to the infinity past of these models.

For the non-adiabatic vacuum computed in Sec.~\ref{subsec:V}, looking at Eq.~\eqref{qseries}, we can see that its series expansion only produces small $q_{n>0}$ for $\lambda < 2 \Rightarrow  w < 1/3$, explaining the upper limit we imposed to $\lambda$. Moreover, for $\lambda = 1 + 3w$ with $0<\vert w\vert\ll1$, the $q_0$ term~\eqref{plq0q1} and, consequently, the VDC given in Eqs.~\eqref{super-Hubbleftest2} and~\eqref{super-Hubbleptest2} diverge very slowly for positive $w$ or converge to zero for negative $w$. In a geometric point-of-view, the VDC in the original frame, given by Eq.~\eqref{nonVDCfull}, tends asymptotically to the ``null''-vector $\vl_2^+$ if $w>0$, and to the origin $\vg_0$ if $w<0$. For positive $w$, this does not mean that the vacuum is not stable. All solutions starting close to the non-adiabatic VDC become closer and closer as we move back in time. That is, the VDC itself is valid everywhere but in the infinite past. We can evaluate the power spectra to understand the differences between these two scenarios. Inspecting Eqs.~\eqref{super-Hubbleftest2} and~\eqref{super-Hubbleptest2} we note that for $0<w\ll1$ the power-spectrum for $\phi$ ($\propto\vert\phi\vert^2\propto T^{-6w}$) diverges slowly while for $\Pi_\phi$ tends to a constant. Now, if $\phi$ diverges, one should account for its back-reaction in the background model. In addition, when $\phi$ is a first-order perturbation, the perturbative approach is unstable in this limit since $\phi$ grows unbounded. Nevertheless, both power-spectra tend to be constants for $-2/3<w<0$. Thus the model in this scenario is completely stable, at least mode-wise.

Furthermore, it also takes longer for the power-spectrum to diverge in the positive case as $w \to 0$ (the case $w=0$ excluded, one cannot have $\sigma_0=0$). Therefore, the optimum situation for maximum stability of the power-spectra is $|3w| \ll 1$. As it is well known, observations indicate that dark matter should satisfy $c_s^2 \ll 1$ and $|w| \ll 1$ (but they cannot be null as well). That may indicate a deep connection between the cosmological constant and dark matter: one can obtain an arbitrary long-lived vacuum state with non-divergent power spectra in the far past of such models only when we consider both components at once. There is also the interesting possibility of considering self-interacting dark-matter components. These have their equation of state modified by bulk-viscosity that can lead to a slightly negative $w$ and consequently to stable models.

The consequences of this finding may be far-reaching. To the best of our knowledge, it is the first time one gets a connection between the cosmological constant and a fluid with the properties of dark matter, two independent and unrelated components of the standard cosmological model. They cooperate to allow the construction of an asymptotic stable quantum vacuum in a sense discussed above. This fact may shed some light on the search of physically meaningful stable vacua in string theory. One usually searches in the direction of a pure cosmological constant, or inflaton fields with slow-roll potentials, without success (see, for example~\cite{Palti2019}). Perhaps one should change to search in the direction of a positive cosmological constant together with a fundamental field suitable to describe dark matter, as K-essence scalar fields, which, as far as we know, were never investigated in this framework. Maybe a positive cosmological constant and dark matter are two fundamental ingredients of gravitation that cannot be studied separately, they seems to be consistent only when combined. It is maybe an interesting avenue to follow.

In upcoming work, we will use the stable quantum vacuum obtained above as the initial quantum state of quantum cosmological perturbations evolving in bouncing models with a positive cosmological constant and dark matter to evaluate their amplitude and spectra.

We will also enlarge the above framework to multiple fields and place it into an extended and more fundamental mathematical framework.

\acknowledgments

SDPV acknowledges the support of CNPq of Brazil under grant PQ-II 316734/2021-7. NPN acknowledges the support of CNPq of Brazil under grant PQ-IB 310121/2021-3. 

\appendix

\section{Phase space representation}
\label{app:PSR}

We can write the product between two vectors $\BRi_a$ and $\BSi_a$ in $\mathbb{V}_{\HFi,t_0}$as $\BRi_a^*\SM^{ab}\BSi_b.$ We will raise and lower indices using the matrix $\SM_{ab}$ and its inverse through $\BRi^a \equiv \SM^{ab}\BRi_b$. From this definition, we have the following properties
\begin{align}
	\BRi_a &= \SM_{ab}\BRi^b,  &\BRi_a^* = \BRi^{b*}\SM_{ba} &,\nonumber \\
	\BRi^a &= \SM^{ab}\BRi_b, &\BRi^{a*} = \BRi_b^*\SM^{ba} &, \label{properties:basis} \\
	\BRi_a^*\BSi^a &= \BRi^{a*}\BSi_a. && \nonumber
\end{align}
If a given vector has a positive norm, i.e., $\BRi_a^*\SM^{ab}\BRi_b > 0$, its complex conjugate has a negative norm $(\BRi_a^*)^*\SM^{ab}\BRi_b^* = -\BRi_b^*\SM^{ba}\BRi_a < 0$. Therefore, it is convenient to choose the basis such that the vector $\BRi_a$ always has positive norm, and its complex conjugate a negative one. Using a normalized vector, $\BRi^{a*}\BRi_a = 1$, we define the projectors
\begin{equation}\label{eq:project}
	\begin{split}
		\Prp_a{}^b \equiv \BRi_a\BRi^{b*}, \qquad \Prp_a{}^{b*} \equiv \BRi^*_a\BRi^b = \SM^{bc}\Prp_c{}^d\SM_{da}.
	\end{split}
\end{equation}
It is easy to show that
\begin{equation}\label{eq:def:complet}
	\begin{split}
		\delta_a{}^b &= \Prp_a{}^b + \Prp_a{}^{b*}, \qquad \SM_{ab} = \Prp_{ab} - \Prp_{ab}^*,
	\end{split}
\end{equation}
where
\begin{equation}\label{defprp}
	\begin{split}
		\Prp_{ab} &= \Prp_a{}^c\SM_{cb} = \BRi_a\BRi_b^*.
	\end{split}
\end{equation}
The expressions above are valid for any unitary vector. This implies that, for
any basis, the imaginary part of the projector $\Prp_{ab}$ is the symplectic
matrix $\SM_{ab}$, and the real part of $\Prp_a{}^b$ is the identity.

We can now define the operator $\CS_a{}^b$ as,
\begin{equation}\label{eq:def:CS}
	\CS_a{}^b \equiv \left(\Prp_a{}^b - \Prp_a{}^{b*}\right), \qquad \CS_{ab} = \left(\Prp_{ab} + \Prp_{ab}^*\right),
\end{equation}
and, conversely,
\begin{equation}\label{def:P}
	\Prp_a{}^b = \frac{\CS_a{}^b + \delta_a{}^b}{2}.
\end{equation}
Here we should note that $\CS_{ab}$ is a real matrix while $\CS_a{}^b$ is purely
imaginary. Hence, the real operator $M_a{}^b$ defined below satisfies,
\begin{equation}\label{complex}
M_a{}^b\equiv -\ci\CS_a{}^b,\qquad	M_a{}^c M_c{}^b = -\delta_a{}^b,
\end{equation}
and defines a complex structure in the phase space. Here we call the attention to the fact that any normalized phase vector $\BRi_a$ leads to a linear structure $M_a{}^c$. However, any other phase vector that differs by a simple phase leads to the same complex structure, i.e., given a phase vector $\BSi_a = e^{\ci\theta}\BRi_a$, the projectors are the same
$$\Prp_a{}^b[\BSi] = \BSi_a\BSi^{a*} = \BRi_a\BRi^{a*} = \Prp_a{}^b[\BRi],$$
and consequently the complex structures are also equal. We discuss this property further in the section below.

\subsection{Reduced phase space equivalence}
\label{app:Mv}

The reduced phase space $\PSVr$ consists of equivalence classes of normalized complex phase vectors $[\BRi_a]$. There is an one-to-one correspondence between the points in $\PSVr$ and the space of linear complex structure $\PSVM$ (the space of $2\times2$ real matrices $M_a{}^b$ satisfying Eq.~\eqref{complex}). First, we show that Eq.~\eqref{eq:def:CS} provides a map $\mu : \PSVr \to \PSVM$,
$$\mu([\BRi])_a{}^b = -\ci\left(\BRi_a\BRi^{b*} - \BRi_a^*\BRi^{b}\right).$$
It is evident that the map is independent of the representative of $[\BRi_a]$,
i.e., any phase vector $\BSi_a \in  [\BRi_a]$ is taken to the same matrix,
\begin{equation}\label{eqrs}
	\BRi_a\BRi^{b*} - \BRi_a^*\BRi^{b} = \BSi_a\BSi^{b*} - \BSi_a^*\BSi^{b}.
\end{equation}
Given any two points in $\PSVr$, $[\BRi]$ and $[\BSi]$ that are taken to
the same matrix $\mu([\BRi])_a{}^b = \mu([\BSi])_a{}^b$, we can contract
Eq.~\eqref{eqrs} with $\BRi^{a*}\BRi_b$ and $\BRi^{a}\BRi_b$ to obtain
\begin{align}
	\vert \BRi^{a}\BSi_a^*\vert^2 +  \vert \BRi^{a}\BSi_a\vert^2 &= 1, \\
	\BRi^a\BSi_a \BRi^b\BSi_b^* &= 0.
\end{align}
There are two possibilities to solve the second equation above, $\BRi^a\BSi_a =
0$ or $\BRi^a\BSi_a^* = 0$. However, if $\BRi^a\BSi_a^* = 0$ then
Eq.~\eqref{eqrs} shows that $\BSi_a = -\BSi_b\BRi^{b}\BRi_a^*$ which is
inconsistent with the normalization condition
$$\BSi_a^*\SM^{ab}\BSi_b = \BRi_a\SM^{ab}\BRi_b^* = -1 \neq 1.$$
Hence, the only possibility is $\BRi^a\BSi_a = 0$ and consequently $\BSi_a = \BSi_b\BRi^{b*}\BRi_a$ where $\vert\BSi_b\BRi^{b*}\vert = 1$, implying that $\BSi_a \sim \BRi_a$. This amounts to show that the map $\mu$ is injective.

Second, we show that the map is surjective. Given any matrix $M_a{}^b$ in $\PSVM$, its eigenvalues must be $\pm\ci$, i.e.,
$$M_a{}^b\BRi_b = \lambda \BRi_a, \quad\Rightarrow\quad -\BRi_a=\lambda^2\BRi_a.$$
As a consequence, since $M_a{}^b$ is real its non-null eigenvectors must be complex. Hence, given any real phase vector $\BVi_a$, the vector $M_a{}^b\BVi_b$ must be linearly independent of it, and it is always possible to build a complex eigenvector with eigenvalue $-\ci$ of $M_a{}^b$ as
\begin{equation}
	\BRi_a \propto  \BVi_a +\ci M_a{}^b\BVi_b,
\end{equation}
which can then be normalized. It is clear that $\BRi_a^*$ is also an eigenvector, but with eigenvalue $+\ci$. This shows explicitly that given any real matrix $M_a{}^b \in \PSVM$ we can define an eigenvector as above. Furthermore, clearly $\BRi_a$ and $\BRi_a^*$ are linearly independent and form a basis in our two-dimensional complex phase space. For these reasons we can write
\begin{equation}\label{Mfromfield}
M_a{}^b = -\ci\left(\BRi_a\BRi^{b*} - \BRi_a^*\BRi^{b}\right),
\end{equation}
showing that $\mu$ is surjective.

\subsection{The $(X, U)$ parametrization}

In the following, we build an explicit representation for the matrices in
$\PSVM$, given the real basis $\BRi^{(i)a} \equiv \delta^{ia}$, $i=
1,\,2$. The action of $M_a{}^b$ on this basis is
\begin{align}
\BRi^{(1)a}M_a{}^b &= X\BRi^{(1)b} + U\BRi^{(2)b},\\ 
\BRi^{(2)a}M_a{}^b &= l_1\BRi^{(1)b} + l_2\BRi^{(2)b},
\end{align}
where $U$ and $l_1$ must be different from zero (otherwise $\BRi^{(i)}_b$ would be
eigenvectors). Applying the same matrix on the expressions above results in
\begin{align}
	-\BRi^{(1)a} &= \left(U l_1 + X^2\right)\BRi^{(1)a} + U\left(X + l_2\right)\BRi^{(2)a}, \\
	-\BRi^{(2)a} &= l_1\left(X+l_2\right)\BRi^{(1)a}+ \left(l_1 U + l_2^2\right)\BRi^{(2)a},
\end{align}
and therefore, $l_1 = -(1+X^2) / U$ and $l_2 = -X$. This means that we can
parametrize the linear complex structure matrix as
\begin{equation}\label{param:XU}
	M_a{}^b \doteq \left(\begin{array}{cc} X & U \\ -{(1+X^2)}/{U} & -X \end{array} \right).
\end{equation}
This parametrization also has a direct interpretation in terms of $\CS_{ab}$ (see Eq.~\eqref{eq:def:CS})
\begin{equation}
\label{jab}
	\CS_{ab} = \ci M_a{}^c\SM_{cb} \doteq \left(\begin{array}{cc} U & -X \\ -X & {(1+X^2)}/{U} \end{array} \right),
\end{equation}
thus,
\begin{align}\label{rel1}
	\CS_{11} &= U = 2\vert \BRi_1\vert^2, \\
	\label{rel2}
	\CS_{12} &= -X = \BRi_1\BRi_2^* + \BRi_1^*\BRi_2, \\
	\label{rel3}
	\CS_{22} &= \frac{1+X^2}{U} = 2\vert \BRi_2\vert^2.
\end{align}
Equation~\eqref{twop} then shows that $U/2$ is just the power-spectrum of
$\hat{\chi}_1 = \hat\mf$, $(1+\chi^2)/(2U)$ the power-spectrum of $\hat{\chi}_2 =
\hat{\Pi}_{\mf}$.  In addition, the product of the variances is just
\begin{equation}
	\braketOP{0_\BRi}{\hat{\mf}^2}{0_\BRi}\braketOP{0_\BRi}{\hat{\Pi}^2_\mf}{0_\BRi} = \frac{1+\chi^2}{4}.
\end{equation}
Hence, $\chi$ measures how far the vacuum $\ket{0_\BRi}$ is from saturating the
uncertainty principle. Since $U \neq 0$, Eq.~\eqref{rel1} shows that
$U\in\mathbb{R}_{>0}$ while $\chi \in\mathbb{R}$. Of course, we could add a
disconnected branch to $\PSVM$ with negative $U$. However, this is not
necessary in this setting and we will consider $\PSVM$ always restricted to
$U>0$.

\subsection{The $(\alpha,\,\gamma)$ parametrization}
\label{app:alphabeta}

Another useful parametrization for the matrix $M_a{}^b$ is $\chi = \sh\alpha$
and $U=\ch(\alpha)e^{-\gamma}$, i.e.,
\begin{equation}\label{eqM}
	M_a{}^b \doteq \left(\begin{array}{cc} \sh\alpha & \ch\alpha\,e^{-\gamma} \\ -{\ch\alpha\,e^{+\gamma}} & -\sh\alpha \end{array} \right).
\end{equation}
Using this parametrization we have $(\alpha,\,\gamma)\in\mathbb{R}^2$, providing
a one-to-one mapping between $\mathbb{R}^2$ and $\PSVM$. Next, the eigenvectors
have a simple form in this parametrization, namely,\footnote{Naturally, there is a particular choice of phase made in this expression.}
\begin{equation}\label{eivenV}
	\BS_a \doteq \left[\frac{e^{-\frac{\gamma}{2}}}{2}\left(\ci e^{-\frac{\alpha}{2}}-e^{\frac{\alpha}{2}}\right),\;\frac{e^{\frac{\gamma}{2}}}{2}\left(\ci e^{-\frac{\alpha}{2}}+e^{\frac{\alpha}{2}}\right)\right].
\end{equation}

Given two points in $\PSVM$, $(\alpha_1,\, \gamma_1)$ and $(\alpha_2,\,
\gamma_2)$, the Bogoliubov coefficients	can be readily computed in this parametrization
through Eq.~\eqref{eq:def:alpha:beta} ,
\begin{align}
	\alpha_{2,1} &= \ch\Delta\alpha\ch\Delta\gamma + \ci \sh\bar\alpha \sh\Delta\gamma, \\
	\beta_{2,1} &= \sh\Delta\alpha\ch\Delta\gamma +\ci \ch\bar\alpha\sh\Delta\gamma,
\end{align}
where $\Delta\gamma \equiv (\gamma_1-\gamma_2)/2$, $\Delta\alpha \equiv
(\alpha_1-\alpha_2)/2$ and $\bar\alpha \equiv (\alpha_1+\alpha_2)/2$.

\subsection{Matrix notation}
\label{app:matrix}

The matrix notation is useful for the calculation related to the time evolution of the linear complex structures. In this appendix we summarize the rules to go from index notation to matrix notation. First, the vector indices are lowered and raised following the rules~\eqref{properties:basis}, that is, they are lowered and raised by the left, and complex conjugated vectors by the right. Second order tensors will follow the pattern in Eq.~\eqref{eq:project}, i.e., the left/right index is raised/lowered by the left/right.

The product $\BRi_a^*\SM^{ab}\BSi_a$ can be written as
\begin{equation}\label{prod}
	\langle \BRi, \BSi\rangle \equiv \BRi_a^*\SM^{ab}\BSi_a.
\end{equation}
Given a linear operator $L_a{}^b$, its action on a vector is simply $(L\BRi)_a = L_a{}^b\BRi_b$, hence, the adjoint can be defined as
\begin{equation}\label{adjoint}
\begin{split}
	\langle\BSi,L\BRi\rangle &= \BSi_a^*\SM^{ab} L_b{}^c\BRi_c = (L^\dagger_a{}^bq_b)^*\SM^{ac}\BRi_c\\
	 &= \langle L^\dagger\BSi,\BRi\rangle,
\end{split}
\end{equation}
where we have, as usual,
\begin{equation}
	L^{\mathrm{T} ab} \equiv L^{ba},\qquad L^{\dagger ab} \equiv L^{\mathrm{T}ab*} = L^{ba*}.
\end{equation}
From the definitions above, we obtain
\begin{align}\label{def:dagger}
	L_a^{\dagger}{}^{b} &= \left(\SM_{ac}L_d{}^{c}\SM^{db}\right)^*,\\
	L_a^{\mathrm{T}}{}^{b} &= -\SM_{ac}L_d{}^{c}\SM^{db}.
\end{align}
Note that, $L_a^\dagger{}^b = L_a{}^b$ iff $L^{\dagger ab} = L^{ab}$, and analogously for $L^\mathrm{T}_a{}^b$. The 2x2 identity matrix is denoted by
$I$. Generally we will denote by $L$ the matrix whose elements are given by $L_a{}^b$, the components of the tensor with the left index down and right index up. Therefore, $(L J)_a{}^b \equiv L_a{}^cJ_c{}^b$. It is worth noting that the definition of ${}^\dagger$ above in matrix notation is
\begin{equation}
L^\dagger \equiv \SM\mathrm{T}(L^*)\SM,
\end{equation}
where T is the matrix transpose. Using these definitions the usual property
$(LJ)^\dagger = J^\dagger L^\dagger$ still holds.

\subsection{Canonical transformations}
\label{app:ct}

The action~\eqref{eq:action:new} can be transformed by any canonical
transformation. However, to keep the equations of motion linear in the fields it
is necessary to restrict to a smaller group that acts linearly on the phase
space and leave the symplectic form invariant. This group is the symplectic
group $\mathrm{Sp(2;\mathbb{R})}$. The group is defined over the field of reals
so the transformations keep the fields Hermitian. Given a canonical
transformation $C_a{}^b$ it is clear that
$$(C_a{}^b\BRi_b)^* \SM^{ac}C_c{}^d\BSi_d = \BRi_a\SM^{ab}\BSi_b.$$
In matrix notation this is the same as
$$  \langle C\BRi, C\BSi\rangle =  \langle C^\dagger C\BRi,\BSi\rangle = \langle\BRi,\BSi\rangle.$$
It is easy to check that $C^\dagger C = I = C
	C^\dagger$ using the definition of ${}^\dagger$ in Eq.~\eqref{def:dagger}.
Moreover, under linear canonical transformation a linear transformation $L$ transforms as
\begin{equation}
L\to CLC^\dagger.
\end{equation}
The algebra of $\mathrm{Sp(2;\mathbb{R})}$ can be written in terms of three generators, namely
\begin{equation}\label{g0-g1-g2}
	\vg_1 \doteq \left( \begin{array}{cc} 1 & 0 \\ 0 & -1\end{array}\!\right), \; \vg_2 \doteq \left( \begin{array}{cc} 0 & 1 \\ 1 & 0\end{array}\!\right), \; \vg_0 \doteq \left( \begin{array}{cc} 0 & 1 \\ -1 & 0\end{array}\!\right).
\end{equation}
Applying the definition of ${}^\dagger$ one can check that $\vg_A^\dagger = -\vg_A$ and the traces $\mathrm{Tr}(\vg_A) = 0$, with the capital Latin indices $A\in\{0,1,2\}$. Furthermore, one can check that these elements, with the addition of the identity matrix $I$ form a bi-dimensional Clifford algebra with
\begin{equation}
	\vg_i\cdot \vg_j = \delta_{ij}\vI, \quad \vg_1\wedge \vg_2 = \vg_0,
\end{equation}
where $i=1,2$ and we defined the operations 
\begin{equation}
\va \cdot \vb \equiv \frac{\va\vb+\vb\va}{2}, \quad \va \wedge \vb = \frac{\va\vb-\vb\va}{2}.
\end{equation}
where $\va$ and $\vb$ are any linear combination of $\vg_A$. The Clifford product is just the matrix multiplication, i.e.,
\begin{equation}
\va\vb = \va\cdot \vb + \va\wedge \vb.
\end{equation}
A useful property is
\begin{equation}
\vg_a\wedge\vg_b = -\epsilon_{ABD} \eta^{DC} \vg_C,
\end{equation}
where  $\epsilon_{ABD}$ is the totally anti-symmetric symbol with $\epsilon_{012}=1$.
It is clear that we can use both the matrix representation forming linear combinations  of $\vI,\;\vg_0,\;\vg_1,\;\vg_2$ or the Clifford algebra considering linear combinations of $1,\;\vg_1,\;\vg_2,\;\vg_1\wedge\vg_2$, which we call multi-vectors.  For simplicity we are going to use the Clifford representation.

General multi-vectors can be written as $\vv=\vv^A\vg_A$. They have an induced internal product given by $\vg_A \cdot \vg_B = \eta_{AB}$, where $\eta_{AB} \doteq \mathrm{diag}(-1,1,1)$. In other words the multi-vectors $\vv$ form a three dimensional vector space with a 1+2 Lorentzian metric. In the Clifford algebra representation it is clear that the dot product of two vectors $\vv = \vv^A\vg_A$ and $\vp=\vp^A\vg_A$ is a scalar, that is 
\begin{equation}
\vv \cdot \vp = \vv^A \vp^B\vg_A\cdot\vg_B = \vv^A \vp^B\eta_{AB}.
\end{equation}
The same computation using the matrix representation would have an identity matrix $\vI$ multiplying the right-hand-side. In order to obtain a scalar, since the traces of $\vg_A$ are all zero and $\mathrm{Tr}(\vI)=2$, one defines the dot product in the matrix representation as
\begin{equation}
\vv \cdot \vp = \frac{\mathrm{Tr}\left(\vv\vp\right)}{2}.
\end{equation}

The square of a multi-vector is an scalar in the Clifford algebra, thus,
$$\vv^2 = \vv \cdot \vv = \vv^A\vv^B\eta_{AB}.$$
For this reason,  the exponential map of an element of the symplectic group algebra can be written as
\begin{equation}\label{exp:v}
	e^\vv = \left\{\begin{array}{ll}
		\cosh(\sqrt{\vv^2})+\frac{\vv}{\sqrt{\vv^2}}\sinh(\sqrt{\vv^2}) & \vv^2>0,\\
		\cos(\sqrt{-\vv^2})+\frac{\vv}{\sqrt{-\vv^2}}\sin(\sqrt{-\vv^2}) & \vv^2<0, \\
		1 + \vv & \vv^2=0.
	\end{array}\right.,
\end{equation}
Applying the ${}^\dagger$ operator to the exponential map results in
$(e^\vv)^\dagger = e^{-\vv}$, therefore, as expected $(e^{\vv})^\dagger e^\vv = 1 =
e^\vv(e^{\vv})^\dagger$. Notice that a linear complex structure $M$
is also an element of this algebra, that is,
\begin{align}
	\label{M-g}
M &=  \vg_0\left(\vg_2 \sinh\alpha  + e^{\vg_1\gamma}\cosh\alpha\right), \\ \nonumber
	&=\cosh\alpha\cosh\gamma\;\vg_0+\sinh\alpha\;\vg_1-\cosh\alpha\sinh\gamma\;\vg_2,
\end{align}
where the parametrization~\eqref{eqM} was used. From the Lorentzian space point-of-view, $M$ is a ``time-like'' vector ($M^2=-1$) with positive time component, $-M\cdot \vg_0 = \cosh\gamma\cosh\alpha > 0$. Hence, the reduced phase space $\PSVM$ is also a subspace of the algebra. Note also that there are simple relations between $M$ and the fields, i.e.,
\begin{equation}\label{Mtofields}
	\begin{split}
M &= \left(\vert r_1\vert^2+\vert r_2\vert^2\right)\vg_0+\left(\vert r_1\vert^2-\vert r_2\vert^2\right)\vg_2\\
&-\left(r_1r_2^*+r_1^*r_2\right)\vg_1,
\end{split}
\end{equation}
for a $M$ defined by Eq.~\eqref{Mfromfield}. Moreover, it is easier to relate the fields with the two null vectors,
\begin{equation}\label{defvl2}
\vl_{2}^\pm \equiv \frac{\vg_2 \pm \vg_0}{2},\qquad \left(\vl_{2}^\pm\right)^2=0,
\end{equation}
that is
\begin{equation}\label{Mtofieldssol}
\begin{split}
   \vert r_1\vert^2 &= M\cdot\vl_2^-,\quad \vert r_2\vert^2 = -M\cdot\vl_2^+,\\
   r_1r_2^*+r_1^*r_2 &= -M\cdot\vg_1.
\end{split}
\end{equation}

Finally, a general Hamiltonian tensor~\eqref{defN} has its associated matrix $N$ represented by
\begin{align}\label{HN}
N &= \left(\frac{1}{m}+m\nu^2\right)\frac{\vg_0}{2}+h\vg_1  + \left(\frac{1}{m}-m\nu^2\right) \frac{\vg_2}{2}, \nonumber \\
	&= \nu\cosh\xi\; \vg_0  + h\; \vg_1 - \nu\sinh\xi\; \vg_2.
\end{align}
This vector has the norm $N^2 = -(\nu^2-h^2)$. If the Hamiltonian is positive definite, then $N$ is ``time''-like.

\subsection{Clifford algebra and $\mathbb{H}^2$}
\label{app:hyspace}

In this subsection we summarize some relations between the Clifford algebra
products and the hyperbolic distances in $\mathbb{H}^2$. For more details and
proofs see~\cite{Ratcliffe2008}. Let $\vv$ and $\vp$ designate two normalized ``time''-like Clifford vectors. The hyperbolic distance between the two points in
$\mathbb{H}^2$ pointed by them is
\begin{equation}\label{def:eta}
	\begin{split}
	d(\vv,\vp) &\equiv \cosh^{-1}\left(-\vv\cdot \vp\right), \\ 
	\vv\cdot \vp &= -\cosh d(\vv,\vp).
	\end{split}
\end{equation}
The second equality above is just the expression of the product in the Clifford
algebra in terms of the distance $d(\vv,\vp)$. One can also compute the modulus
of the vector $\vv \wedge \vp$, first note that this product has a simple relation
with the Lorentzian cross product $\otimes$ (as defined in~\cite{Ratcliffe2008}),
\begin{equation}
	\vv\wedge \vp = - \vv\otimes \vp.
\end{equation}
Thus, the square of this product is simply
\begin{equation}
	\left(\vv\wedge \vp\right)^2 = \left(\vv\otimes \vp\right)^2 = \sinh^2 d(\vv,\vp).
\end{equation}
This shows that the product of two normalized ``time''-like vectors is a
``space''-like vector with norm $\sinh d(\vv,\vp)$. Furthermore, we define the
modulus of a ``space''-like vector $\vl$,
\begin{equation}
	\vert \vl \vert \equiv \sqrt{\vl^2}.
\end{equation}
Consequently, $\vert \vv\wedge \vp\vert = \sinh d(\vv,\vp)$.

It is easy to show that for any two non-null vectors $\vl$ and $\vm$, we have
\begin{equation}
	(\vl\wedge \vm)\cdot \vm = 0 = (\vl\wedge \vm)\cdot \vl.
\end{equation}
This shows that $\vv\wedge \vp$ is always orthogonal to both $\vv$ and $\vp$. Now, given $\vv$ and $\vp$ as above, we can compute the perpendicular part of $\vv$ with respect to $\vp$,
\begin{equation}
	\vv\perp \vp  \equiv \vv+(\vv\cdot \vp)\vp = (\vp\wedge \vv)\vp = (\vp\wedge \vv)\wedge \vp.
\end{equation}
The norm of this vector is simply
\begin{equation}
	(\vv \perp \vp)^2 = (\vp\wedge \vv)^2 = \sinh^2 d(\vv,\vp),
\end{equation}
showing also that $ \vv\perp \vp $ is ``space''-like and has the same length as
$\vp\wedge \vv$. One can also check that
\begin{align}
	(\vv\perp \vp)\wedge \vp &= (\vv\perp \vp) \vp = \vv\wedge \vp, \\
	(\vv\perp \vp)\cdot \vp &= 0, \\
	(\vv\perp \vp)\cdot (\vv\wedge \vp) &= 0, \\
	(\vv\perp \vp)\wedge (\vv\wedge \vp) &= \vp\sinh^2 d(\vv,\vp).
\end{align}
The results above show that given $\vv$ and $\vp$ we can build an orthogonal frame
$\vp$, $\vv\perp \vp$ and $\vv\wedge \vp$. We denote with an overbar a ``space''-like
normalized vector, i.e.,
\begin{equation}\label{normalized}
	\widebar{\vl} = \frac{\vl}{\vert \vl\vert}.
\end{equation}
Finally, in general we can write,
\begin{equation}\label{decompvp}
	\vv = \cosh d(\vv,\vp)\,\vp + \sinh d(\vv,\vp) \widebar{\vv\perp \vp}.
\end{equation}

\section{The Hamiltonian tensor}
\label{app:HT}

In general, the Hamiltonian tensor introduced in Eq.~\eqref{eq1} can be written
as
\begin{equation}\label{Horig}
\gH^{ab} \doteq \left( \begin{array}{cc} m\nu^2 & h \\ h & \frac{1}{m} \end{array} \right),
\end{equation}
Even though, $\gH_a{}^b$ is self-adjoint, the phase space is not a proper
Hilbert space (note that any real phase vector has zero norm). For this reason,
we write the Hamiltonian tensor in the form~\eqref{Horig}. Using the fact
that $\gH^{ab}$ is a symmetric matrix, it is clear that it has a well defined
eigenvalue problem,
\begin{equation}
\delta_{ab}\gH^{bc}\BRi^{(i)}_c = \lambda_i\BRi^{(i)}_a,
\end{equation}
where $i=1,\,2$, $\lambda_{i}$ are the real eigenvalues and $\BRi^{(i)}_a$ the
associated real eigenvectors. There are three possibilities depending on the
signs of $\lambda_{i}$:\footnote{Naturally, we have
	all the other cases that can be obtained by multiplying the whole tensor
	$\gH^{ab}$ by $-1$, however, they are equivalent to their counterparts.}
\begin{enumerate}
	\item $\lambda_{1} > 0$  and $\lambda_{2} > 0$, positive definite Hamiltonian,
	\item $\lambda_{1} > 0$  and $\lambda_{2} < 0$,
	\item $\lambda_{1} > 0$  and $\lambda_{2} = 0$.
\end{enumerate}
In all cases above we have two real eigenvectors $\BRi^{(i)}_a$ that are
orthonormal, i.e., $\bar{\BRi}^{(i)a}\BRi^{(j)}_a = \delta^{ij}$, where we
introduced the over bar $\bar{\BRi}^{(i)a} \equiv \delta^{ab}\BRi^{(i)}_b$ to
differentiate from the index raised with $\SM^{ab}$. An eigenvector $\BSi_a$ of
$\gH_a{}^b$ with eigenvalue $\sigma$, if it exists, can be written was a linear combination of
$\BRi^{(i)}_a$,
\begin{equation}
\BSi_a = \sum_{i=1}^2c_i\BRi^{(i)}_a,
\end{equation}
where $c_i$ are arbitrary constants. Applying the Hamiltonian tensor $\gH_a{}^b$
on this vector results in
\begin{align*}
&\gH_a{}^b\BSi_b  = \sigma \BSi_a, \\
&\SM_{ab}\left(c_1\lambda_1\bar\BRi^{(1)b} + c_2\lambda_2\bar\BRi^{(2)b}\right)=\sigma\left(c_1\BRi^{(1)}_a + c_2\BRi^{(2)}_a\right).
\end{align*}
Contracting the second equation above with $\bar\BRi^{(1)a}$ and
$\bar\BRi^{(2)a}$, results respectively in
\begin{align}\label{c1c2}
c_2\lambda_2\bar\BRi^{(1)a}\SM_{ab}\bar\BRi^{(2)b} &= \sigma c_1, \\
c_1\lambda_1\bar\BRi^{(2)a}\SM_{ab}\bar\BRi^{(1)b} &= \sigma c_2.
\end{align}
Which can be combined as
\begin{equation*}
\sigma\left(c_1^2\lambda_1+c_2^2\lambda_2\right) = 0.
\end{equation*}
The solutions for $\sigma \neq 0$ are $c_1 = \pm \ci c_2
\sqrt{\lambda_2/\lambda_1}$. The tensor $\SM_{ab}$ is an anti-symmetric tensor
in a bi-dimensional space and form therefore a one-dimensional space. Thus, it
must be proportional to $\BRi_a^{(1)}\BRi_b^{(2)}-\BRi_a^{(2)}\BRi_b^{(1)}$. The
proportionality constant can be obtained using that $\SM_{ab}\SM^{bc} =
\delta_a{}^c$. Performing this calculation we obtain that
$\bar\BRi^{(1)a}\SM_{ab}\bar\BRi^{(2)b} = \pm\ci$, using the freedom to multiply
the eigenvectors $\BRi_a^{(i)}$ by $-1$, we choose
$\bar\BRi^{(1)a}\SM_{ab}\bar\BRi^{(2)b} = \ci$ . Substituting these results back
into Eq.~\eqref{c1c2}, results in
\begin{equation}\label{solEV}
\BSi_a = c_2\left(\pm\ci\sqrt{\frac{\lambda_2}{\lambda_1}}\BRi^{(1)}_a + \BRi^{(2)}_a\right).
\end{equation}
That is, both $\BSi_a$ and $\BSi_a^*$ are eigenvectors with eigenvalues $\pm\sqrt{\lambda_1\lambda_2}$ respectively. For this reason, we choose $\sigma$ as the positive eigenvalue, 
$$\sigma = \sqrt{\lambda_1\lambda_2}.$$
In the case 1,  we can normalize $\BSi_a$ as
\begin{equation}
\BSi_a^*\SM^{ab}\BSi_b = 2\vert c_2\vert^2\sqrt{\frac{\lambda_2}{\lambda_1}} = 1, \quad c_2 = \sqrt{\frac12\sqrt{\frac{\lambda_1}{\lambda_2}}},
\end{equation}
where we had to choose the plus sign in Eq.~\eqref{solEV}, otherwise the norm
would be negative. In cases 2 and 3, the eigenvector $\BSi_a$ is real (apart
from a overall phase) and cannot be normalized. This means that $\gH_a{}^b$
specify normalizable eigenvectors iff $\gH^{ab}$ is positive definite.

The Sylvester's criterion for positive definite operators provide the conditions
under which an operator is positive definite. In the case of the real matrix
$\gH^{ab}$, we obtain
\begin{equation}\label{syl}
m > 0,\quad \nu^2-h^2>0,  \quad\mathrm{and}\quad \nu^2 > 0.
\end{equation}
If these conditions are satisfied, we can reparametrize $h_\nu \equiv
\tanh^{-1}(h/\nu) $ and $\xi \equiv\ln(m\nu)$. Under these conditions
it is straightforward to show that the real eigenvalues are
\begin{align}
\lambda_1 & = \nu\left(\cosh\xi + \sqrt{\sinh^2\xi+\tanh^2 h_\nu}\right),\\
\lambda_2  &= \nu\left(\cosh\xi - \sqrt{\sinh^2\xi+\tanh^2 h_\nu}\right), \\
\sigma &= \frac{\nu}{\cosh h_\nu}.
\end{align}
Since the tensor $\gH_a{}^b$ is purely imaginary, we define another real tensor
\begin{equation}\label{defN}
\begin{split}
N_a{}^b&\equiv -\ci\gH_a{}^b,\\
&\doteq \sigma\left( \begin{array}{cc} \sinh h_\nu & \cosh h_\nu e^{-\xi} \\ -\cosh h_\nu e^{\xi} & -\sinh h_\nu \end{array} \right).
\end{split}
\end{equation}
Comparing to Eq.~\eqref{eqM} it is easy to see that $N_a{}^b/\sigma$ is actually a linear complex structure. That means that, if $\gH^{ab}$ is a positive definite matrix, it induces the linear complex structure $N_a{}^b/\sigma$ above. As it happens with $M_a{}^b$, $N_a{}^b/\sigma$ has eigenvectors equivalent to Eq.~\eqref{eivenV}, namely
\begin{equation}\label{eivenVN}
\BS_a \doteq \left[\frac{e^{-\frac{\xi}{2}}}{2}\left(\ci e^{-\frac{h_\nu}{2}}-e^{\frac{h_\nu}{2}}\right),\;\frac{e^{\frac{\xi}{2}}}{2}\left(\ci e^{-\frac{h_\nu}{2}}+e^{\frac{h_\nu}{2}}\right)\right].
\end{equation}

\section{Adiabatic frames and adiabatic propagator}
\label{app:adiab}

The computation of the adiabatic frames can be greatly simplified by making a first canonical transformation in order to eliminate $h$ from the Hamiltonian~\eqref{Horig}. Note that if the off-diagonal term $h$ is present, the Hamiltonian vector has non-null components on both directions $\vg_1$ and $\vg_2$ (see Eq.~\eqref{HN}). It is easy to see that the canonical transformation
\begin{equation}
	\label{remove-h}
	C_h = e^{\frac{h m}{2}(\vg_2-\vg_0)},
\end{equation}
takes $\vn$ as in Eq.~\eqref{HN} to
\begin{align}
	\label{bar-n}
	\vn' &= C_h \vn C_h^\dagger + \dot{C}_hC_h^\dagger \nonumber \\
	&= \left(\frac{1}{m}+m{\sigma}^{\prime2}\right) \frac{\vg_0}{2} +  \left(\frac{1}{m}-m{\sigma}^{\prime2} \right) \frac{\vg_2}{2},
\end{align}
where $$\sigma^{\prime2} \equiv \nu^2 - h^2-\dot{h}-h\frac{\dot{m}}{m},$$
leading to the new equations of motion
\begin{equation}
	\dot{\vm}' = 2\; \vn'\wedge \vm' = 2{\sigma'}\; \vu'\wedge \vm',
\end{equation}
where
\begin{equation}
	\vu'\equiv \frac{\vn'}{\sigma'} = \cosh\xi'\, \vg_0-\sinh\xi'\, \vg_2, \quad \xi' \equiv \ln (m\sigma').
\end{equation}
Hence, this canonical transformation removes the $\vg_1$ term from $\vn$, or equivalently, it removes the off-diagonal term $h$ in the Hamiltonian~\eqref{Horig}, and redefines the frequency $\sigma^2 = \nu^2 - h^2$. The new frequency $\bar{\sigma}$ is still positive definite, since $\sigma^2 = \nu^2-h^2>0$ (see~\eqref{syl}), while the derivatives are first order in the adiabatic approximation, and they must be small when compared with the other terms, by assumption. From now on, we will work in this frame, and, for notation simplicity, we will omit the primes.

This parametrization makes it clear that
\begin{equation}
	\label{u0}
	\vu=\cosh\xi\, \vg_0-\sinh\xi\, \vg_2
\end{equation}
is a boost applied to $\vg_0$, i.e.,
\begin{equation}
	\label{boost1}
	\vu = e^{\frac{-\xi}{2} \vg_0\wedge \vg_2} \vg_0 e^{+\frac{\xi}{2} \vg_0 \wedge\vg_2}= e^{-\frac{\xi}{2} \vg_1} \vg_0 e^{\frac{\xi}{2} \vg_1}.
\end{equation}
Hence, the parameters of the canonical transformation given by Eq.~\eqref{CTug0} transforming $\vu$ to $\vg_0$ can be easily read from Eq.~\eqref{boost1}, $e^{\frac{\xi}{2} \vg_1} \vu e^{\frac{\xi}{2} \vg_1}=\vg_0$, leading to $d_\vu = \vert\xi\vert$ and $\widebar{\vu\wedge \vg_0} = \mathrm{sign}(\xi)\vg_0 \vg_2 = \mathrm{sign}(\xi)\vg_1$. As a consequence, our first canonical transformation reads,
\begin{equation}
	\label{C1}
	\vm^{(1)} \equiv C_{0}\vm C_{0}^\dagger = e^{\frac{\xi}{2} \vg_0\vg_2}\vm e^{-\frac{\xi}{2} \vg_0\vg_2}.
\end{equation}
Using Eq.~\eqref{du1} to calculate the Hamiltonian correction, one gets
\begin{align}\label{adiabN1}
	\vn^{(1)} &= \sigma\left(\vg_0 - F_1\vg_1\right) ,  & F_1 &\equiv -\frac{\dot{\xi}}{2\sigma}.
\end{align}
The normalized Hamiltonian now reads
\begin{equation}
	\label{u1}
	\vu^{(1)} = \cosh\xi_1\,\vg_0-\sinh\xi_1\vg_1,
\end{equation}
while the new frame ${}^{(1)}$ parameters are,
\begin{equation}
	\sigma_1 \equiv \sigma\sqrt{1-F_1^2}, \qquad\xi_1 \equiv \tanh^{-1}\left(F_1\right).
\end{equation}
The equations of motion are given by
\begin{equation}
	\dot{\vm}^{(1)} = 2{{\sigma}_1}\; {\vu}^{(1)}\wedge {\vm}^{(1)}.
\end{equation}

Comparing Eqs.\eqref{u0} and \eqref{u1}, one can see that the normalized Hamiltonian vector at first order $\vu^{(1)}$ has the same form as the zero order $\vu$, with the ``space''-like direction rotated to $\vg_1$. It is therefore clear that the Hamiltonian in this frame is also a boost from $\vg_0$, parametrized by $\xi_1$,
\begin{equation}
	\vu^{(1)} = e^{-\frac{\xi_1}{2} \vg_0\vg_1} \vg_0 e^{+\frac{\xi_1}{2} \vg_0\vg_1} = e^{\frac{\xi_1}{2} \vg_2} \vg_0 e^{-\frac{\xi_1}{2} \vg_2},
\end{equation}
Hence, again, the above inverse canonical transformation takes $\vu^{(1)}$ to $\vg_0$. This induces the second order frame, where
\begin{align}
	\vn^{(2)} &= \sigma_1\left(\vg_0 - F_2\vg_2\right), & F_2 &\equiv +\frac{\dot{\xi}_1}{2\sigma_1},
\end{align}
whereas the normalized Hamiltonian, frequency and boost parameters are
\begin{align}
	\vu^{(2)} &= \cosh\xi_2\,\vg_0-\sinh\xi_2\vg_2, \\
	\sigma_2 &= \sigma_1\sqrt{1-F_2^2}, \qquad
	\xi_2 \equiv \tanh^{-1}\left(F_2\right),
\end{align}
while the equations of motion read
\begin{equation}
	\dot{\vm}^{(2)} = 2{\sigma_2}\; {\vu}^{(2)}\wedge {\vm}^{(2)}.
\end{equation}

Repeating the process in order to go to the frame ${}^{(n)}$ from the frame ${}^{(n-1)}$ leads to the boost parameters
\begin{equation}\label{boostparams}
	\begin{split}
		F_n &= (-1)^n\frac{\dot{\xi}_{n-1}}{2\sigma_{n-1}}, \qquad \xi_{n} = \tanh^{-1}\left(F_n\right),\\
		\sigma_{n} &= \sigma_{n-1}\sqrt{1-F_{n}^2},
	\end{split}
\end{equation}
where the initial functions of this recurrence relations are
\begin{equation}
	\sigma_{0} \equiv \sigma, \qquad \xi_{0} \equiv \xi.
\end{equation}
The first boost, parametrized by $\xi$ is not necessarily small, $\xi$ can have any value. Nonetheless, all the subsequent boosts have small parameters $\xi_{n} \ll 1$ for $n>0$. Note that these transformations are finite and exact. Moreover, in each frame the term $F_n$ yields corrections of adiabatic order $\bigO{n}$ to the Hamiltonian multi-vector. 

We can now compute the approximate propagator in a frame ${}^{(n)}$, which satisfies
\begin{equation}
\label{prop-ad-apc}
\dot {P}^{(n)}(t) =N^{(n)}(t)P^{(n)}(t),
\end{equation}
and the Hamiltonian multi-vector reads 
\begin{equation}
	\vn^{(n)}(t)  = \sigma_{n-1}(t)\left(\vg_{0} - F_n(t)\vg_n\right).
\end{equation}

Eq.~\eqref{prop-ad-apc} for $P^{(n)}$ can be rewritten as
\begin{equation}
\frac{\dd}{\dd t}\left(e^{-\vg_0\Delta\tau}P^{(n)}\right) =-\sigma_{n-1}F_{n}e^{-\vg_0\Delta\tau}\vg_nP^{(n)}(t),
\end{equation}
where 
\begin{equation}
\Delta\tau(t)\equiv\exp\left(\int_{t_0}^t \sigma_{n-1}(t')\dd t' \right).	
\end{equation}

Introducing the rotating propagator $P^{(n)}_r$ and the spatial vector $\va^{(n)}$ as
\begin{align}
P^{(n)}_r&\equiv e^{-\vg_0\Delta\tau}P^{(n)},& \va^{(n)} &\equiv e^{-\vg_0\Delta\tau}\vg_ne^{\vg_0\Delta\tau},
\end{align}
the  equation of motion for $P_r^{(n)}$ reads
\begin{equation}
\dot{P}^{(n)}_r	 = -\sigma_{n-1}F_{n} \va^{(n)}P^{(n)}_r.
\end{equation}
The integral version of this equation can be readily obtained as
\begin{equation}
\label{rot-prop}
P^{(n)}_r(\tau,\tau_0) = 1 - \int_{\tau_0}^\tau F_{n}(\tau')\va^{(n)}(\tau')P^{(n)}_r(\tau',\tau_0)\dd \tau'.
\end{equation} 
Note that, although the vector $\va^{(n)}$ depends on time, it has always modulus one, $\left(\va^{(n)}\right)^2 = 1$. Therefore, the integrand in the right-hand-side of Eq.~\eqref{rot-prop} is of adiabatic order $\bigO{n}$. Substituting the expression for the propagator back into its own expression leads to
\begin{equation}
P_r^{(n)}(\tau,\tau_0) = 1 -p^{(n)}(\tau,\tau_0) +\bigO{2n},
\end{equation} 
where 
\begin{equation}
p^{(n)}(\tau,\tau_0) \equiv \int_{\tau_0}^\tau F_n(\tau')\va^{(n)}(\tau')\dd \tau'.
\end{equation}
Naturally, up to this same order we can use the following propagator
\begin{equation}
P^{(n)}_r(\tau,\tau_0) = e^{-p^{(n)}(\tau,\tau_0)}.
\end{equation}
Going back to the non-rotating frame the propagator is
\begin{equation}
P^{(n)}(\tau,\tau_0) = e^{\vg_0\Delta\tau}e^{-p^{(n)}(\tau,\tau_0)}.
\end{equation}
It is more convenient to rewrite the propagator changing the order of the two terms, that is,
\begin{equation}
\label{propagator(n)}
P^{(n)}(\tau,\tau_0) = e^{-p_r^{(n)}(\tau,\tau_0)}e^{\vg_0\Delta\tau},
\end{equation}
where we defined a new multi-vector
\begin{equation}
p_r^{(n)}(\tau,\tau_0) =e^{\vg_0\Delta\tau}p^{(n)}(\tau,\tau_0)e^{-\vg_0\Delta\tau}.
\end{equation}
Before applying this propagator to an initial condition, it is worth studying the behavior of the spatial vector $p_r^{(n)}(\tau,\tau_0)$. First we rewrite it as
\begin{equation}\label{barp}
p_r^{(n)}(\tau,\tau_0) =\vg_n\int_{\tau_0}^\tau F_n(\tau')e^{-2(\tau-\tau')\vg_0}\dd \tau' .
\end{equation}
If $n<N_\mathrm{max}$, we can integrate by parts to obtain
\begin{equation}
	\begin{split}
p_r^{(n)}(\tau,\tau_0) &= \vg_0\wedge\vg_{n}\left[\left( \frac{F_n(\tau)}{2} -\frac{F_n(\tau_0)}{2}e^{-2\Delta\tau\vg_0}\right)\right.\\
&\left.-\int_{\tau_0}^\tau \frac{\dot{F}_n(\tau')}{2\sigma_{n-1}}e^{-2(\tau-\tau')\vg_0}\dd \tau'\right].
\end{split}
\end{equation}
The second term is of adiabatic order $n+1$, and we can neglect it. Moreover, the functions $F_n(\tau)$ usually have a well-defined limit where the adiabatic corrections are asymptotically zero. Choosing the time $\tau_0$ to match this limit we get
\begin{equation}
\label{propagator(n)-part}
p_r^{(n)}(\tau) \approx \frac{F_n(\tau)}{2}\vg_0\wedge\vg_{n}.
\end{equation}
Inserting Eq.~\eqref{propagator(n)-part} back into Eq.~\eqref{propagator(n)}, we finally get the propagator
\begin{equation}\label{proporder}
P^{(n)}(\tau,\tau_0) \approx e^{-\vg_0\wedge\vg_{n}\frac{F_n(\tau)}{2}}e^{\Delta\tau\vg_0}.
\end{equation} 
The expression above shows that the time evolution (up to order $n$) is the rotation about the origin $\vg_0$ followed by a boost in the $\vg_0\wedge\vg_{n}$ direction. This is just the indication that there is another adiabatic frame of order ${}^{(n+1)}$ (see Eqs.~\eqref{boostparams} and~\eqref{Cj}).

Now, if $n=N_\mathrm{max}$ then the integration by parts is not justified and we need another way to estimate Eq.~\eqref{barp}. Nevertheless, using that
\begin{align*}
\vert F_n(\tau')\cos \left[2(\tau-\tau')\right]\vert \leq \vert F_n(\tau')\vert, \\
\vert F_n(\tau')\sin \left[2(\tau-\tau')\right]\vert \leq \vert F_n(\tau')\vert,
\end{align*}
and Eq.~\eqref{boostparams}, we notice that the components of the multi-vector $p_r$ are always smaller than
\begin{equation}
\int_{\tau_0}^\tau \vert F_n(\tau')\vert \dd \tau' = \int_{\tau_0}^\tau \left\vert \frac12 \frac{\dd{\xi}_{n-1}}{\dd\tau'}\right\vert \dd \tau' \leq \frac{\vert\xi_{n-1}\vert}{2},
\end{equation}
assuming that $\xi_{n-1}$ is largest at $\tau$. This shows that the final boost does not have a well-defined direction, but it is still  generated by a parameter smaller than the adiabatic order $n-1$. In this way, we complete the proof that propagator \eqref{proporder} is the one to be used up to order $n$.

Finally, the points on frame ${}^{(n)}$ can be mapped back to the original frame by applying the inverse canonical transformations, that is,
\begin{align}
	\vv(t) &= C_{0}^\dagger ... C_{n-1}^\dagger \vv^{(n)}(t) C_{n-1}... C_{0},
\end{align}
where a general canonical transformation $C_{j}$  reads
\begin{align}
	\label{Cj}
	C_{j} &= e^{\frac{\xi_{j}}{2} \vg_0\vg_{j}}, & \vg_j &\equiv \left\{\begin{array}{ccc} \vg_1 & j & \mathrm{odd} \\ \vg_2 & j & \mathrm{even} \end{array} \right..
\end{align}
Two multi-vectors in two adjacent adiabatic orders are related by
\begin{equation}
V^{(n-1)} = C_{n-1}^\dagger \vv^{(n)}(t) C_{n-1}.
\end{equation}
Writing the multi-vector in terms of its components, i.e., $V^{(n)} = V^{(n)A}\vg_A$ result in the following relation between the components,
\begin{equation}
\begin{split}
V^{(n-1)0} &= V^{(n)0}\cosh\xi^{(n-1)} - V^{(n)2}\sinh\xi^{(n-1)},\\
V^{(n-1)1} &= V^{(n)1}, \\ 
V^{(n-1)2} &= V^{(n)2}\cosh\xi^{(n-1)}-V^{(n)0}\sinh\xi^{(n-1)}, 
\end{split}
\end{equation}
where we assume $n-1$ even (there is a similar relation for $n-1$ odd). The expression above is simply the result of a boost applied to a ``time''-like vector $V^{(n)}$. The final relation between the frame $(n)$ and the original frame depend on $n$ noncollinear boosts. It is straightforward to compose a sequence of boosts in a single boost  and rotation where the last is known as Wigner rotation (see, for example~\cite{Ferraro1999}).  Here, we are interested in the relation between frames up to a given adiabatic order. For example, using the parametrization described in~\ref{app:alphabeta}, we can write $(\alpha, \beta)$ in terms of $(\alpha_2, \beta_2)$ defined in frame ${}^{(2)}$ as
\begin{align}
\alpha &= \alpha_2 - \xi_1\cosh\gamma_2 - \frac{\xi_1^2\sinh\gamma_2\tanh\alpha_2}{2}+\bigO{3}, \\
\nonumber \gamma &= \gamma_2 + \xi + \xi_1\sinh\gamma_2\tanh\alpha_2\\
&\xi_2+\frac{\xi_1^2}{4}\left[\sinh\left(2\gamma_2\right)\left(1-\frac{2}{\cosh^2\alpha_2}\right)\right]+\bigO{3}.
\end{align}
These expressions are valid transformations from any point in the frame ${}^{(2)}$ to the original frame. In particular, the origin at ${}^{(2)}$, that is $V^{(2)}=\vg_0$ is mapped to
\begin{align}
	\alpha &= - \xi_1+\bigO{3}, \\
	\gamma &= \xi + \xi_2+\bigO{3},
\end{align}
which are the usual well known adiabatic parameters.

\bibliographystyle{apsrev}

\bibliography{references}

\end{document}